\documentclass[11pt,a4paper]{article}
\pdfoutput=1
\usepackage{jheppub}
\usepackage{amsfonts}
\usepackage{amsmath}
\usepackage{amssymb}
\usepackage{cancel}
\usepackage{color}
\usepackage[table,dvipsnames]{xcolor}
\usepackage{graphicx}
\usepackage{hyperref}
\usepackage{mathtools}
\usepackage{pifont}
\usepackage{placeins}
\usepackage{url}
\usepackage{multirow}
\usepackage{slashed}
\usepackage{makecell}
\usepackage{xspace}
\usepackage{float}
\usepackage{bm}
\usepackage{tikz}

\renewcommand{\arraystretch}{1.4}

\newcommand{\blue}[1]{{\color{blue} #1}}

\newcommand{\qt}[1]{``#1''}

\newcommand{\sigbr}{$\sigma \cdot {\cal B}$\xspace}
\newcommand{\HEPfit}{\texttt{HEPfit}\xspace}
\newcommand{\cp}{$\mathcal{CP}$\xspace}
\newcommand{\rn}{{\tt r}}
\newcommand{\Mn}{{\tt M}}
\newcommand{\R}{\text{Re}}
\newcommand{\I}{\text{Im}}

\newcolumntype{P}[1]{>{\centering\arraybackslash}p{#1}}
\newcolumntype{Q}{>{\columncolor{cms}}c}
\newcolumntype{R}{>{\columncolor{atlas}}c}
\newcolumntype{S}{>{\columncolor{lep}}c}

\colorlet{atlas}{Apricot!70!White}
\colorlet{lep}{Lavender!50!White}

\definecolor{color8TeV}{HTML}{ddddff}
\definecolor{color13TeV}{HTML}{eeee99}
\definecolor{cms}{HTML}{9DCEBF}

\allowdisplaybreaks

\begin{document}

\title{Light scalars within the $\mathcal{CP}$-conserving Aligned-two-Higgs-doublet model}

\author[a]{Antonio M. Coutinho,}
\emailAdd{antonio.coutinho@ific.uv.es}

\author[a]{Anirban Karan,}
\emailAdd{kanirban@ific.uv.es} 

\author[b]{V\'ictor Miralles,}
\emailAdd{victor.miralles@manchester.ac.uk}

\author[a]{Antonio Pich}
\emailAdd{antonio.pich@ific.uv.es}

\affiliation[a]{Instituto de F\'isica Corpuscular, Universitat de València -- CSIC, Parque Cient\'ifico, Catedr\'atico Jos\'e Beltr\'an 2, E-46980 Paterna, Spain}

\affiliation[b]{Department of Physics and Astronomy, University of Manchester, Oxford Road, Manchester M13 9PL, United
Kingdom}

\abstract{  
In this article we study the possibility that neutral and charged scalars lighter than the 125~GeV Higgs boson might exist within the framework of the $\mathcal{CP}$-conserving Aligned-two-Higgs-doublet model. Depending on which new scalar (scalars) is (are) light, seven different scenarios may be considered. Using the open-source code {\tt HEPfit}, which relies on Bayesian statistics, we perform global fits for all seven light-mass scenarios. The constraints arising from vacuum stability, perturbativity, electroweak precision observables, flavour observables, Higgs signal strengths, and direct-detection results at the LEP and the LHC are taken into account. Reinterpreted data from slepton searches are considered too. It turns out that the seven scenarios contain sizeable regions of their parameter space compatible with all current data. 
Although not included in the global fits, the possible implications of $(g-2)_\mu$ are also addressed.
}

\preprint{ }

\maketitle

\section{Introduction}
\label{sec:Intro}

The observation of the Higgs boson at the LHC in 2012~\cite{ATLAS:2012yve, CMS:2012qbp} firmly cemented the Standard Model (SM) as \textit{the theory of almost everything}~\cite{Oerter:2006}, our currently accepted picture of elementary particles and their interactions. Still, as the `almost' term denotes, it is widely known that the SM possesses a set of shortcomings that will require the addition of new degrees-of-freedom, which might address all or a few of the issues, while leaving intact what is already known. Among the candidates to build extensions to the SM are scalar fields transforming as doublets of $SU(2)_L$. These are well motivated, if not for anything else, by the fact that fermions have shown Nature's propensity to bear multiple families of fields: the minimal SM contains a single doublet, but without particular laws prohibiting their existence, it stands to reason that Nature may have produced more than one, following the old adage that \qt{everything not forbidden is compulsory}~\cite{Gell-Mann:1956iqa, White:1958}. Moreover, additional doublets with the same quantum numbers as the SM Higgs leave the tree-level value of the $\rho$ parameter equal to unity~\cite{Diaz-Cruz:2003kcx} and, while extra generations of quarks and leptons are strongly constrained from the Higgs production cross-section at the LHC~\cite{ATLAS:2016neq,Ilisie:2012cc}, unitarity-triangle data~\cite{UTfit:2022hsi} or $Z$-boson branching ratios~\cite{ALEPH:2005ab}, augmenting the SM with scalar doublets circumvents such stringent bounds.

The simplest of such models with multiple doublets is the two-Higgs-doublet model (2HDM)~\cite{Branco:2011iw, Gunion:1989we}, where only one copy is added, and every term that contains the SM Higgs is duplicated. Beyond the three Goldstone bosons, which become the longitudinal polarisations of $Z$ and $W^\pm$ and give mass to these gauge bosons, the 2HDM comes with one charged and three neutral scalars. This enlarged scalar spectrum brings many interesting possibilities with it, such as
new sources of \cp violation~\cite{Lee:1973iz, Wu:1994ja, Botella:1994cs, Gunion:2005ja, Grzadkowski:2013rza, Keus:2015hva, Chen:2017com, Iguro:2019zlc},
the smallness of neutrino masses~\cite{Ma:2006km, Hirsch:2013ola},
dark matter candidates~\cite{Barbieri:2006dq, LopezHonorez:2006gr, Belyaev:2016lok, Tsai:2019eqi},
particles with axion-like behaviour~\cite{Wilczek:1977pj, Kim:1986ax, Celis:2014zaa, Allison:2014hna, Espriu:2015mfa, Anuar:2024myn},
electroweak baryogenesis~\cite{Turok:1990zg, Cline:2011mm, Fuyuto:2015jha, Dorsch:2016nrg, Basler:2021kgq, Enomoto:2022rrl, Goncalves:2023svb}
and the full stability of the vacuum until the Planck scale~\cite{Das:2015mwa, Ferreira:2015rha, Staub:2017ktc, Branchina:2018qlf, Schuh:2018hig, Ferreira:2019bij}
which the SM does not seem able to provide~\cite{Degrassi:2012ry, Buttazzo:2013uya, Bednyakov:2015sca}.
It also comes with its own set of problems, namely the fact that the 2HDM, in its most generic form, introduces tree-level flavour-changing neutral currents (FCNC) that might contribute to certain phenomena beyond what current experimental limits allow. To tackle these pernicious FCNC, the usual implementations of the model impose specific discrete $Z_2$ symmetries which ensure that each type of right-handed fermions couples with only one of the doublets, thus getting rid of the tree-level flavour-changing interactions altogether~\cite{Glashow:1976nt, Paschos:1976ay}. An alternative, less stark solution to the FCNC issue is the alignment of Yukawa couplings in flavour space, so that the diagonalisation of mass terms also trims the non-diagonal elements of the matrices associated with the new neutral scalars. This is the very idea behind the Aligned-two-Higgs-doublet model (A2HDM)~\cite{Pich:2009sp, Pich:2010ic, Penuelas:2017ikk}, whose more generic framework adds complexity and offers a richer phenomenology with respect to the $Z_2$-symmetric realisations of the model --- both at colliders~\cite{Abbas:2015cua, Celis:2013rcs, Celis:2013ixa} and low-energy flavour experiments~\cite{Jung:2010ik, Jung:2010ab, Jung:2012vu}. If complex, the parameters that govern the alignment can be new sources of \cp violation, beyond those usually on offer for 2HDMs: the parameters of the scalar potential not invariant under rephasings, and the Cabibbo–Kobayashi–Maskawa (CKM) matrix~\cite{Pich:2009sp}. While the alignment condition is scale-dependent and radiative corrections introduce deviations from it at higher loops~\cite{Ferreira:2010xe}, it has been shown that this proportionality can naturally stem from simple discrete symmetries and be stable under renormalisation-group evolution~\cite{Botella:2015yfa}. Indeed, in general, the flavour violation is minimal~\cite{Chivukula:1987py, DAmbrosio:2002vsn, Manohar:2006ga, Buras:2010mh} and FCNC are not only absent at tree-level but also strongly suppressed at higher perturbative orders, ensuring that they do not impact Higgs phenomenology~\cite{Pich:2009sp, Pich:2010ic, Jung:2010ik, Braeuninger:2010td, Bijnens:2011gd, Li:2014fea, Abbas:2015cua, Penuelas:2017ikk, Gori:2017qwg}.

Explorations of 2HDMs allowing for scalars lighter than 125~GeV have been published in the literature over the years, most recently in Refs.~\cite{Iguro:2023jju, Mondal:2023wib, Fu:2023sng, Azevedo:2023zkg, Belyaev:2023xnv, Arhrib:2023apw, Ma:2024deu, Arhrib:2024sfg, Biekotter:2024ykp, Liu:2024azc, Benbrik:2024ptw, Doff:2024hid, Das:2024ekt, Khanna:2024bah, Dong:2024ipo, Banik:2024ugs, Li:2024kpd, Arhrib:2024hed}. These recent probes have, for the most part, been focused on possible hints from data for a scalar with a mass around 95~GeV~\cite{LEPWorkingGroupforHiggsbosonsearches:2003ing, CMS:2018cyk, CMS:2024yhz, ATLAS:2024bjr} and how it could exist within a 2HDM, mainly in its version with a $Z_2$ symmetry. Concerning the A2HDM, its low-mass region has also been analysed in the past, yet either in scans over limited sections of parameter space or dedicated to tests of a specific set of observables~\cite{Jung:2010ik, Akeroyd:2012yg, Jung:2012vu, Altmannshofer:2012ar, Celis:2013rcs, Celis:2013ixa, Li:2014fea, Ilisie:2014hea, Enomoto:2015wbn, Han:2015yys, Cherchiglia:2016eui, Cherchiglia:2017uwv}. A global fit to the A2HDM with no further sources of \cp violation beyond the CKM matrix was first performed in Ref.~\cite{Eberhardt:2020dat} and subsequently updated in Refs.~\cite{Karan:2023kyj, Karan:2023xze}. Both analyses took a Bayesian approach to statistical inference, powered by the Markov Chain Monte Carlo (MCMC) framework of the publicly available software \HEPfit~\cite{DeBlas:2019ehy}, yet, in both cases, the spectrum of scalar masses was mostly limited to the heavy scenario, where all new Higgs bosons are assumed to be heavier than the one already observed at the LHC. In this work, we set out to perform, for the first time, a global analysis of the \cp-conserving A2HDM, once more with the aid of \HEPfit, and following on from preliminary results we have already included in two proceedings papers~\cite{Karan:2024kgr, Coutinho:2024vzm}. With respect to the two previous global fits, this inspection of an enlarged range of masses required the addition of CMS and ATLAS probes of light scalars to the card of LHC data used in the code; additionally, due to their relevance for the regions below 120~GeV for neutral scalars, and below 100~GeV for charged scalars, we have included the bounds from direct searches at the LEP. Elsewhere, we have brought any colliders' or low-energy experiments' data to their state-of-the-art values, in case they have been updated since the fit of Ref.~\cite{Karan:2023kyj}. We pay particular attention to the anomalous magnetic moment of the muon, which has seen significant developments in recent years. Given the uncertainty that still surrounds its SM value, owing to the existing discrepancies between different methods used in its estimation, we do not include this observable in our main fits. Nevertheless, we devote one section to the examination of the possible influence that $(g-2)_\mu$ could have
on our fits, something we also do to the direct searches for supersymmetric (SUSY) particles at the LEP and the LHC that may be reinterpreted as searches for bosons from a generic extended scalar sector.

This paper is organised as follows: in section \ref{sec:A2HDM} we present a brief overview of the model; section \ref{sec:Const} contains an exposition of the constraints that can be used in fits to the A2HDM; in section \ref{sec:Setup}, we explain the software and statistical setup used to perform our global analyses, whose results are then shown and discussed in section \ref{sec:Plots}; in section \ref{sec:gm2}, we address how different theory predictions of $(g-2)_\mu$ would affect our fits; finally, our main conclusions are given in section \ref{sec:Concl}.

\section{The Aligned-two-Higgs-doublet model}
\label{sec:A2HDM}

The A2HDM is one of the simplest extensions of the SM, as it involves extending the scalar sector by introducing only one additional complex $SU(2)_L$ doublet with a hypercharge of $1/2$, exactly like the SM Higgs boson. In general, after electroweak symmetry breaking, the neutral components of both scalar doublets could acquire a vacuum expectation value (vev). However, they can always be redefined through a rotation under the $SU(2)_L \otimes U(1)_Y$ group without loss of generality. This rotation can be chosen such that only one of the two scalars acquires a non-zero vev. This basis is commonly referred to as the Higgs basis, in which the scalar fields after electroweak symmetry breaking can be expressed as
\begin{equation}
\Phi_1=\frac{1}{\sqrt 2}\begin{pmatrix}
\sqrt 2\;G^+\\
v+S_1+i\, G^0
\end{pmatrix}\, ,\qquad\qquad \Phi_2=\frac{1}{\sqrt 2}\begin{pmatrix}
\sqrt 2\;H^+\\
S_2+i\, S_3
\end{pmatrix}\, ,
\end{equation} 
where $\Phi_1$ gets the vev  $v=246$~GeV. The components $G^\pm$ and $G^0$ can be identified as the Goldstone bosons which provide the masses to the $W^\pm$ and $Z$ bosons. Consequently, the remaining physical spectrum consists of a single charged scalar $H^\pm$, two \cp-even neutral scalars $S_{1,2}$, and one \cp-odd neutral scalar $S_3$. The neutral scalars $S_i$ do not necessarily correspond to the mass eigenstates, since a non-diagonal mass matrix is generated by the most general scalar potential allowed by the SM gauge symmetries:
\begin{align}
\label{eq:pot}
V&=\mu_1\,\Phi_1^\dagger \Phi_1+\mu_2\,\Phi_2^\dagger \Phi_2+ \Big[\mu_3\,\Phi_1^\dagger \Phi_2+\mathrm{h.c.}\Big]+\frac{\lambda_1}{2}\,(\Phi_1^\dagger \Phi_1)^2+\frac{\lambda_2}{2}\,(\Phi_2^\dagger \Phi_2)^2+\lambda_3\,(\Phi_1^\dagger \Phi_1)(\Phi_2^\dagger \Phi_2)\nonumber\\
&+\lambda_4\,(\Phi_1^\dagger \Phi_2)(\Phi_2^\dagger \Phi_1)+\Big[\Big(\frac{\lambda_5}{2}\,\Phi_1^\dagger \Phi_2+\lambda_6 \,\Phi_1^\dagger \Phi_1 +\lambda_7 \,\Phi_2^\dagger \Phi_2\Big)(\Phi_1^\dagger \Phi_2)+ \mathrm{h.c.}\Big]\, ,
\end{align}
where $\mu_3,\,\lambda_5,\,\lambda_6$, and $\lambda_7$ are complex parameters and the rest are real. Note that there are fewer degrees of freedom than available variables in this equation since the minimisation condition imposes that 
\begin{equation}
v^2=-\frac{2\mu_1}{\lambda_1}=-\frac{2\mu_3}{\lambda_6}\, .
\end{equation}
Furthermore, by choosing the appropriate phase rotation on $\Phi_2$ one of the complex phases of $\lambda_5,\,\lambda_6$, and $\lambda_7$ can be removed, allowing one of these parameters to be taken as real. As such, a total of 11 degrees of freedom can be counted in the scalar sector alone: $\mu_2,\; v,\; \lambda_{1,2,3,4},\; |\lambda_{5,6,7}|$ and the two relative phases between $\lambda_{5,6,7}$.

The mass terms of this potential can be written as
\begin{equation}
\label{eq:mass_pot}
    V_M= \left(\mu_2+\frac{1}{2}\lambda_3 v^2\right) H^+H^-+\frac{1}{2}\begin{pmatrix} S_1 & S_2 & S_3 \end{pmatrix} \mathcal{M} \begin{pmatrix} S_1 \\ S_2 \\ S_3 \end{pmatrix}\, ,
\end{equation}
with 
\begin{equation}
    \mathcal{M}=\begin{pmatrix} v^2\, \lambda_1  & v^2\, \rm{Re}(\lambda_6) & -v^2\, \rm{Im}(\lambda_6) \\
    v^2\, \rm{Re}(\lambda_6) & \left(\mu_2+\frac{1}{2}v^2\lambda_3\right)+\frac{1}{2}v^2\left(\lambda_4+\rm{Re}(\lambda_5)\right) & -\frac{1}{2}v^2\, \rm{Im}(\lambda_5) \\
    -v^2\, \rm{Im}(\lambda_6) & -\frac{1}{2}v^2\, \rm{Im}(\lambda_5) & \left(\mu_2+\frac{1}{2}v^2\lambda_3\right)+\frac{1}{2}v^2\left(\lambda_4 - \rm{Re}(\lambda_5) \right) \end{pmatrix}\! .
    \label{eq:massmatrix}
\end{equation}
The physical neutral states $h$, $H$, and $A$ would, therefore, be the linear combinations of $S_1$, $S_2$, and $S_3$ that diagonalise this mass matrix. In the \cp-conserving limit of the scalar sector, we can take all the $\lambda_{5,6,7}$ as real and the mass matrix becomes block diagonal. Indeed, the mass eigenstates are, in this case, well-defined \cp states: two of them are \cp-even ($h$, $H$) whilst the other is \cp-odd ($A$). In fact, the \cp-odd scalar is just $S_3$ in this case, since the well-defined \cp states do not mix with each other. In order to diagonalise the mass matrix we only need then to rotate the two \cp-even neutral states,
\begin{equation}
\begin{pmatrix}
h\\H
\end{pmatrix}=\begin{pmatrix}
\cos\tilde{\alpha}& \sin\tilde{\alpha}\\ -\sin\tilde{\alpha}&\cos\tilde{\alpha}
\end{pmatrix}\begin{pmatrix}
S_1\\S_2
\end{pmatrix} 
\, ,
\end{equation}
where the scalar $h$ will always be for us the 125~GeV Higgs boson and the scalar $H$ will be the additional new-physics (NP) scalar. From this relation, and taking into account that this rotation diagonalises Eq.~\eqref{eq:massmatrix} to the masses of $h$ ($M_h$) and $H$ ($M_H$), we obtain the relation
\begin{equation}
\tan\tilde\alpha=\frac{M_h^2 -v^2\,\lambda_1}{v^2\,\lambda_6}=\frac{v^2\,\lambda_6}{v^2\,\lambda_1-M_H^2}\, .
\label{eq:mix_ang_lam}
\end{equation}
Using this relation in combination with Eq.~\eqref{eq:mass_pot}, we can relate some of the parameters of the potential to the physical masses:
\begin{align}
\nonumber  \mu_2=M_{H^\pm}-&\frac{\lambda_3}{2}v^2\, ,\;\; \lambda_1=\frac{M_h^2+M_H^2\tan^2{\tilde{\alpha}}}{v^2(1+\tan^2{\tilde{\alpha}})}\, ,\;\;  \lambda_4=\frac{1}{v^2}\left( M_h^2+M_A^2-2M_{H^\pm}^2+\frac{M_H^2-M_h^2}{1+\tan^2{\tilde{\alpha}}}\right),\\
     &\lambda_5=\frac{1}{v^2}\left( \frac{M_H^2+M_h^2\tan^2{\tilde{\alpha}}}{1+\tan^2{\tilde{\alpha}}}-M_A^2\right)\, ,\;\;\; \lambda_6=\frac{(M_h^2-M_H^2)\tan{\tilde{\alpha}}}{v^2(1+\tan^2{\tilde{\alpha}})}\,.
\end{align}
Therefore, we can choose the following nine independent parameters from the scalar sector of the \cp-conserving A2HDM: \(\{v,\,M_{h},\,M_{H^\pm},\,M_{H},\,M_{A},\,\tilde{\alpha},\,\lambda_2,\,\lambda_3,\,\lambda_7\}\), of which the first two have already been precisely measured.

Besides the self-interactions emerging from the scalar potential, the kinetic terms generate interactions of the additional scalars with the gauge bosons. Since $S_1$ is the only scalar acquiring a vev, it plays the role of the SM Higgs regarding the triple bosonic interactions with two SM gauge bosons. For the mass eigenstates these interactions become
\begin{equation}
g_{hVV}=\cos\tilde{\alpha}\;g_{hVV}^{SM}\, , \qquad\qquad g_{HVV}=-\sin\tilde{\alpha}\;g_{hVV}^{SM}\, , \qquad\qquad g_{AVV}=0\, ,
\label{eq:Higgs_weak_boson_couplings}
\end{equation}
with $VV\equiv W^+W^-, ZZ$.

The gauge symmetries also allow for the emergence of a Yukawa interaction 
\begin{align}
-\mathcal L_Y\, =\, &\bigg(\! 1+\frac{S_1}{v}\!\bigg)\Big\{\bar u_L\,M_u\,u_R+\bar d_L\,M_d\,d_R+\bar l_L\, M_l\,l_R\Big\}\nonumber\\
&+\frac{1}{v}\, (S_2+i S_3) \Big\{\bar u_L\,Y_u\,u_R+\bar d_L\,Y_d\,d_R+\bar l_L\, Y_l\,l_R\Big\}\\
&+\frac{\sqrt 2}{v}\, H^+ \Big\{\bar u_L\,V\,Y_d\,d_R-\bar u_R \,Y_u^\dagger\,V\,d_L+\bar\nu_L\, Y_l\,l_R\Big\} + \mathrm{h.c.}\, ,\nonumber
\end{align}
in the Higgs basis of scalars and mass-basis of fermions, where we have omitted the generation indices, and the subscripts $L$ and $R$ refer to the usual left- and right-handed chiral fields. The matrices $M_f$ ($f\equiv u,\,d,\,l$) are the diagonal mass matrices of quarks and leptons generated by the vev of $\Phi_1$, $Y_f$ are the Yukawa matrices coupled to the doublet with zero-vev $\Phi_2$, and $V$ is the usual CKM matrix. In general, if $Y_f$ are arbitrary $3\times3$ matrices, these Yukawa interactions would generate FCNC at tree level, which, as mentioned in the introduction, are experimentally highly suppressed. In order to avoid these effects, in the A2HDM an alignment condition between $M_f$ and $Y_f$ is imposed in the flavour space~\cite{Pich:2009sp,Pich:2010ic} such that 
\begin{equation}\label{eq:Yalignment}
Y_u=\varsigma^*_u\,M_u \qquad\qquad \text{and} \qquad\qquad Y_{d,l}=\varsigma_{d,l}\,M_{d,l}\, ,
\end{equation}
with $\varsigma_f$, in general, arbitrary complex numbers, yet chosen to be real in our analysis since we assume \cp conservation in the NP throughout this work.

In the scalar mass-eigenstate basis, the interaction with the fermions is given by
\begin{eqnarray}
-\mathcal L_Y &\supset &\bigg(\frac{\sqrt 2}{v}\bigg) H^+\,\Big[\bar u\,\big\{\varsigma_d V M_d \mathcal{P}_R-\varsigma_u M_u^\dagger V\mathcal{P}_L\big\}\, d+\varsigma_l\, \bar \nu M_l \mathcal P_R l\Big] + \mathrm{h.c.}
\nonumber\\ &  & +\;
\sum_{i,f}\;\bigg(\frac{y_f^{\varphi^0_i}}{v}\bigg)\,\varphi^0_i\,\Big[\bar f M_f \mathcal{P}_R f\Big] \, ,
\end{eqnarray}
with $\varphi^0_i$ representing the neutral scalars ($h$, $H$, and $A$) and $\mathcal{P}_{L,R}$ the usual chiral projectors. The couplings $y_f^{\varphi^0_i}$ are related to the parameters defined above by
\begin{align}
&y_{u}^H=-\sin\tilde\alpha+\varsigma_{u}\,\cos \tilde\alpha\, , \qquad\quad y_{u}^h=\cos\tilde\alpha+\varsigma_{u}\,\sin \tilde\alpha\, , \qquad\quad
y_{u}^A= -i\varsigma_{u}\, ,\nonumber\\
&y_{d,l}^H=-\sin\tilde\alpha+\varsigma_{d,l}\,\cos \tilde\alpha\, , \quad\quad\; \;y_{d,l}^h=\cos\tilde\alpha+\varsigma_{d,l}\,\sin \tilde\alpha\, , \quad\quad\;  
y_{d,l}^A= i\varsigma_{d,l}\, .
\label{eq:Higgs_yuk}
\end{align}

As mentioned before, the A2HDM is a more general theoretical framework that includes as particular cases the usual 2HDMs based on discrete $Z_2$ symmetries. To recover them we just need to impose $\mu_3=\lambda_6=\lambda_7=0$ (in the basis where the $Z_2$ symmetry is enforced), along with
\begin{align}
\label{eq:THDM_types}
&\hskip -.1cm\text{Type I:\;\,} \varsigma_{u}=\varsigma_d=\varsigma_l=\cot\beta,\;\;
\text{Type II:\;\,} \varsigma_{u}=-\frac{1}{\varsigma_d}=-\frac{1}{\varsigma_l}=\cot\beta\, ,\;\;
\text{Inert:\;\,}\varsigma_{u}=\varsigma_d=\varsigma_l=0\, ,
\nonumber\\
&\quad\text{Type X:\;\;} \varsigma_{u}=\varsigma_d=-\frac{1}{\varsigma_l}=\cot\beta,
\qquad \text{and}\qquad
\text{Type Y:\;\;} \varsigma_{u}=-\frac{1}{\varsigma_d}=\varsigma_l=\cot\beta \, .
\end{align}
While the imposition of a $Z_2$ symmetry protects against the appearance of FCNC under renormalisation, this protection is in general absent in the A2HDM because higher-order quantum corrections induce a misalignment between $M_f$ and $Y_f$. Nevertheless, it is worth noting once more that the symmetries of the A2HDM Yukawa structure impose strong constraints on potential FCNC arising from this misalignment, rendering their effects numerically negligible~\cite{Pich:2009sp, Pich:2010ic, Jung:2010ik}. 
Even if exact alignment is assumed at a very high-energy scale, such as the Planck scale, the FCNC effects induced by the renormalisation-group running to low energies remain well below current experimental bounds~\cite{Braeuninger:2010td, Bijnens:2011gd, Li:2014fea, Abbas:2015cua, Penuelas:2017ikk, Gori:2017qwg}.

\section{Constraints}
\label{sec:Const}

In the following, we describe in detail all the different theoretical constraints and experimental observables that we have used in our analysis.

\subsection{Vacuum stability}

In order to avoid large values of the scalar potential growing in the negative direction, which would make the model unstable, for any given configuration of the scalar fields, the potential must be bounded from below. Various \textit{necessary} conditions on the parameters $\lambda_i$ for circumventing such large negative potential have recently been discussed in Ref.~\cite{Bahl:2022lio}. The \textit{necessary and sufficient} conditions that ensures the potential to be bounded from below are derived in Refs.~\cite{Ivanov:2006yq, Ivanov:2015nea}. First, one rewrites the scalar potential $V$ (given by Eq.~\eqref{eq:pot}) in the Minkowskian \textit{bilinear formalism}: 
\begin{equation}
\label{eq:sc_pot_2}
     V=-\,\Mn_\mu\,{\rn}^\mu + \frac{1}{2}\,\Lambda^{\mu}_{\phantom{\mu}\nu}\, \rn_\mu\,\rn^\nu\, ,
\end{equation}
where,
\begin{align}
 \Mn_\mu\, &=\, \Big[-\frac{1}{2}(\mu_1+\mu_2),\,-\,\R\,\mu_3, \,\I\,\mu_3,\, -\,\frac{1}{2}(\mu_1-\mu_2)\Big]\, , 
 \nonumber\\  
 \rn^\mu\, &=\, \Big[|\Phi_1|^2+|\Phi_2|^2, \, 2\,\R (\Phi_1^\dagger\Phi_2), \, 2\,\I (\Phi_1^\dagger\Phi_2), \, |\Phi_1|^2-|\Phi_2|^2\Big]\, , \nonumber\\
 \Lambda^{\mu}_{\phantom{\mu}\nu}\, &=\,\frac{1}{2}\,\begin{bmatrix}
     \;\frac{1}{2}(\lambda_1+\lambda_2)+\lambda_3 && \R (\lambda_6+\lambda_7) && -\,\I (\lambda_6+\lambda_7) && \frac{1}{2}(\lambda_1-\lambda_2)\\[-1.5mm]
     - \R (\lambda_6+\lambda_7) && -\lambda_4-\R \lambda_5&& \I \lambda_5 && -\, \R(\lambda_6-\lambda_7)\\[-1.5mm]
     \I (\lambda_6+\lambda_7) && \I \lambda_5&& -\lambda_4+\R \lambda_5 &&  \I(\lambda_6-\lambda_7)\\[-1.5mm]
     -\frac{1}{2}(\lambda_1-\lambda_2)&& -\R (\lambda_6-\lambda_7) && \I (\lambda_6-\lambda_7)&&-\frac{1}{2}(\lambda_1+\lambda_2)+\lambda_3\;\;
 \end{bmatrix}\, .
\end{align}
Depending on the \qt{\textit{timelike}} $(\Lambda_0)$ and \qt{\textit{spacelike}} $(\Lambda_{\{1,2,3\}})$ eigenvalues of the mixed-symmetric matrix $\Lambda^{\mu}_{\;\;\nu}$, the \textit{bounded from below} constraint can be recast in three conditions: a) the tensor $\Lambda^{\mu\nu}$ is diagonalisable by a $SO(1,3)$ transformation, b) all the eigenvalues of $\Lambda^{\mu}_{\;\;\nu}$ are real, and c) $\Lambda_0>0$ with $\Lambda_0>\Lambda_{\{1,2,3\}}$, which is equivalent to the statement that \textit{the tensor $\Lambda^{\mu\nu}$ is positive definite in the forward light-cone}~\cite{Ivanov:2006yq}.

The requirement that the vacuum of the scalar potential should be a stable neutral minimum implies another constraint on the quartic couplings: defining the determinant of the matrix $(\xi\, \mathbb{I}_4-\Lambda^{\mu}_{\;\;\nu})$ as $D=-\prod_{k=0}^3(\xi-\Lambda_k)$, with the Lagrange multiplier $\xi=\frac{M_{H^\pm}^2}{v^2}$, the existence of a global minimum is ensured if a) $D>0$, or b) $D<0$ with $\xi>\Lambda_0$~\cite{Ivanov:2015nea}. Thus, vacuum stability imposes constraints on the mass parameters, quartic couplings, and the mixing angle of the scalars. 

\subsection{Perturbativity}
\label{sec:perturbativity}
\textit{Perturbative unitarity} requires that, at every order of perturbation theory, the scattering amplitudes do not increase monotonically with energy. Thus, imposing this constraint on every $2\to2$ scattering amplitude involving scalars will validate the applicability of the perturbative expansion of the $S$-matrix by restricting the quartic coupling constants $\lambda_i$. Denoting the matrix of tree-level partial wave amplitudes by $\mathbf{a_0}$ and $a_j^0$ its corresponding eigenvalues in the $j^{th}$ partial wave, the conditions for \textit{tree-level unitarity} can be expressed as: 
\begin{equation}
\label{eq:pert}
    (a_j^{0})^2\leq \frac{1}{4} \qquad\text{with}\qquad (\mathbf{a_0})_{i,f}=\frac{1}{16\pi s}\int_{-s}^{0} dt \;\mathcal M_{i\to f}(s,t)\, .
\end{equation}
However, at tree-level, the $S$-wave contribution dominates the scattering amplitude at very high energies, and hence it is enough to consider only the $j=0$ partial wave. The three doubly charged, eight singly-charged, and fourteen neutral two-body scattering states of the scalars can be represented in block-diagonal form as
\begin{eqnarray}
&\mathbf{a_0^{++}}=\frac{1}{16\pi}\,  X_{(1,1)}\; ,\quad 
\mathbf{a_0^+}=\frac{1}{16\pi}\, \text{diag}\, [X_{(0,1)},\, X_{(1,0)},\, X_{(1,1)}]\; ,&\nonumber\\
&\mathbf{a_0^0}=\frac{1}{16\pi}\, \text{diag}\, [X_{(0,0)},\, X_{(0,1)},\, X_{(1,1)},\, X_{(1,1)}]\; ,&  
\end{eqnarray}
where $X_{(Y,I)}$, representing the scattering amplitudes for states with definite hypercharge and weak isospin, are given by~\cite{Ginzburg:2005dt,Bahl:2022lio}:
\begin{alignat}{5}
\label{eq:pert1}
    & X_{(1,0)}=&&\;\lambda_3-\lambda_4\, ,
    && X_{(1,1)}=&&\begin{bmatrix}\lambda_1&&\lambda_5&&\sqrt 2 \lambda_6\\[-1.5mm]
    \lambda_5^*&&\lambda_2&&\sqrt 2 \lambda_7^*\\[-1.5mm]
    \sqrt 2\lambda_6^*&&\sqrt 2 \lambda_7^*&&\lambda_3+\lambda_4\end{bmatrix} , &\nonumber\\
    & X_{(0,1)}=&&\begin{bmatrix}
        \lambda_1&&\lambda_4&&\lambda_6&&\lambda_6^*\\[-1.5mm]   \lambda_4&&\lambda_2&&\lambda_7&&\lambda_7^*\\[-1.5mm]
       \lambda_6^*&&\lambda_7^*&&\lambda_3&&\lambda_5^*\\[-1.5mm]  
       \lambda_6&&\lambda_7&&\lambda_5&&\lambda_3\\[-1.5mm] 
    \end{bmatrix},\quad
    && X_{(0,0)}=&&\begin{bmatrix} 3\lambda_1&&2\lambda_3+\lambda_4&&3\lambda_6&&3\lambda_6^*\\[-1.5mm]   2\lambda_3+\lambda_4&&3\lambda_2&&3\lambda_7&&3\lambda_7^*\\[-1.5mm] 3\lambda_6^*&&3\lambda_7^*&&\lambda_3+2\lambda_4&&3\lambda_5^*\\[-1.5mm] 3\lambda_6&&3\lambda_7&&3\lambda_5&&\lambda_3+2\lambda_4 
    \end{bmatrix}.
\end{alignat}
Hence, \textit{tree-level perturbative unitarity} is guaranteed by demanding the eigenvalues $(e_i)$ of all these $X_{(Y,I)}$ matrices to satisfy:
\begin{equation}
    |e_i| \leq 8\pi\, .
\end{equation}

In the \cp-conserving A2HDM the cubic interactions of the neutral and charged scalars are given by
\begin{equation}
    V\supset v\,H^+H^-\Big[(\lambda_3\,\cos\tilde{\alpha}+\lambda_7\,\sin\tilde{\alpha})\, h+(\lambda_7\,\cos\tilde{\alpha}-\lambda_3\,\sin\tilde{\alpha})\, H\Big]\, .
    \label{eq:pot_charge_neutral}
\end{equation}
Therefore, there is a tree-level cubic interaction between the neutral (\cp-even) and charged scalars. These couplings receive a finite vertex correction from the charged scalars themselves~\cite{Celis:2013rcs}:
\begin{equation}
(\lambda_{\varphi_i^0 H^+ H^-})_{\text{eff}}=\lambda_{\varphi_i^0 H^+ H^-}\left[1+\frac{v^2\, \lambda_{\varphi_i^0 H^+ H^-}^2}{16\pi^2 M_{H^\pm}^2} \mathcal{Z}\left(\frac{M^2_{\varphi_i^0}}{M^2_{H^\pm}}\right)\right]\equiv\lambda_{\varphi_i^0 H^+ H^-}(1+\Delta)\, ,
\end{equation}
with
\begin{equation}
    \mathcal{Z}(X)=\int^1_0\text{d}y\int^{1-y}_0\text{d}z [(y+z)^2+X(1-y-z-yz)]^{-1}\, .
    \label{eq:pert_phiHpHp}
\end{equation}
As suggested in Ref.~\cite{Celis:2013rcs}, we impose the perturbative constraint $\Delta\leq0.5$ to guarantee that quantum corrections do not break the perturbative expansion. Thus, like vacuum stability, perturbativity also puts restrictions on the mass parameters, quartic couplings, and the mixing angle of the scalars.

In order to respect perturbativity in the Yukawa sector, we assume the fermionic couplings of the charged scalar to be smaller than unity, which implies: $|\varsigma_f|< v/(\sqrt 2\, m_f)$.

\subsection{Electroweak precision observables}

The additional scalar particles modify the vacuum polarisation corrections of the electroweak gauge bosons through their contributions to loop diagrams. These loop effects alter the gauge boson propagators and are parametrised by the oblique parameters (also known as Peskin–Takeuchi parameters~\cite{Peskin:1990zt, Peskin:1991sw}) $S$, $T$, and $U$, which encapsulate the deviations caused by NP. In our fit, we only include $S$ and $T$ as observables, assuming $U$ to be negligible --- because the contributions to $U$ are highly suppressed in the A2HDM~\cite{Haber:2010bw}. However, the standard experimental determination of the oblique parameters is obtained from a global electroweak fit~\cite{deBlas:2016ojx,deBlas:2021wap,deBlas:2022hdk}, which also includes the observable \mbox{$R_b\equiv\Gamma (Z\to b\bar b)/\Gamma(Z\to\mathrm{hadrons})$}~\cite{Haber:1999zh,Degrassi:2010ne} as an input. $R_b$ is affected by the additional scalars of the A2HDM. Therefore, using this observable in the determination of the oblique parameters would be inconsistent.
In order to use uncontaminated values for the oblique parameters, we adopt as inputs the results provided in our previous work~\cite{Karan:2023kyj} that were obtained removing $R_b$ from the electroweak fit. For more details, we refer the reader to Appendix B.2 of Ref.~\cite{Karan:2023kyj}. Removing $R_b$ from the oblique-parameter fit is particularly important because we also use $R_b$ as an observable in our analysis, and we need to be sure that we do not double count observables.
These observables provide significant restrictions to the splitting of the masses of the new scalars.

\subsection{Flavour observables}
\label{sec:flavour}

The presence of new scalar particles coupling to quarks and leptons modifies several precisely measured low-energy observables. Some of these processes are used by the CKMfitter~\cite{Charles:2004jd} and UTfit~\cite{UTfit:2006vpt,UTfit:2022hsi} collaborations to determine the flavour structure of the SM, providing precise determinations of the CKM matrix. In our fit, we employ the Wolfenstein parametrisation~\cite{Wolfenstein:1983yz} of the CKM matrix, treating its four free inputs as nuisance parameters to which we assign a Gaussian uncertainty. One must mind, though, that the processes used to determine the CKM parameters should not be contaminated by the presence of NP. Even if the UTfit collaboration provides values for the Wolfenstein parameters removing the loop-level processes~\cite{UTfit:2005lis}, which are the ones mostly affected by NP, the UTfit fit also includes some tree-level processes that receive contributions form the A2HDM scalars. As such, we proceed as in our previous work~\cite{Karan:2023kyj} and repeat the fit of the Wolfenstein parameters, removing from the fit any \blue{observable} that could be contaminated. For more details on this procedure and the final inputs used, we refer the reader to Appendix B.1 of Ref.~\cite{Karan:2023kyj}.

To globally constrain the A2HDM, we take into account all flavour observables that are relevant for \cp-conserving NP. This includes contributions to loop-level processes such as
neutral-meson mixing in the $B_s$ system ($\Delta M_{B_s}$)~\cite{Jung:2010ik,Chang:2015rva},
the weak radiative decay $B \rightarrow X_s \gamma$~\cite{Bobeth:1999ww, Misiak:2006ab, Misiak:2006zs, Jung:2010ik, Jung:2010ab, Jung:2012vu, Hermann:2012fc, Misiak:2015xwa, Misiak:2017woa, Misiak:2020vlo},
and the rare leptonic decay $B_s \rightarrow \mu^+ \mu^-$~\cite{Li:2014fea,Arnan:2017lxi}. 
In addition to these loop-induced processes, we also incorporate relevant tree-level transitions. These include leptonic decays of heavy pseudoscalar mesons such as $B \rightarrow \tau \nu$, $D_{(s)} \rightarrow \mu \nu$, and $D_{(s)} \rightarrow \tau \nu$, as well as ratios of leptonic decays of light pseudoscalar mesons, specifically $\Gamma(K \rightarrow \mu \nu)/\Gamma(\pi \rightarrow \mu \nu)$, and the analogous ratio of tau decays \mbox{$\Gamma(\tau \rightarrow K \nu)/\Gamma(\tau \rightarrow \pi \nu)$}~\cite{Jung:2010ik}. These flavour observables constrain the maximum allowed values for the alignment parameters, depending on the scalar masses, especially the charged Higgs mass.

\subsection{Higgs signal strengths}

The introduction of additional scalar fields significantly influences both the production and decay channels of the observed Higgs boson. The one-loop amplitude of $h\rightarrow\gamma\gamma$ is affected by extra contributions arising from the charged scalar. Moreover, the mixing between the \cp-even scalar states alters the couplings between the SM-like Higgs and the weak gauge bosons, as outlined in Eq.~\eqref{eq:Higgs_weak_boson_couplings}. This modification affects Higgs production via vector-boson fusion (VBF) and its associated production with vector bosons ($Vh$). Similarly, scalar mixing plays a critical role in the Higgs decay to fermions, as highlighted in Eq.~\eqref{eq:Higgs_yuk}. This, in turn, has implications for Higgs production mechanisms such as gluon fusion (ggF) and associated production with top quark-antiquark pairs ($t\bar{t}h$).

Since the discovery of the Higgs boson, its properties have been extensively scrutinised at the LHC. In particular, its most relevant production modes 
(ggF, VBF, $Vh$, and $t\bar{t}h$) and its subsequent decay channels ($c\bar{c},\, b\bar{b},\, \gamma\gamma,\, \mu^+\mu^-,\, \tau^+\tau^-,\, W^+W^-,\, Z\gamma$, and $ ZZ $) have been measured (or bounded) by the ATLAS and CMS collaborations. Both collaborations provide data parametrised in terms of the Higgs signal strengths, which measure the production cross section in a particular production mode times the branching ratio for a given decay channel, in units of the corresponding SM prediction. In this work, we incorporate the complete covariance matrices of signal strength measurements from several key datasets. These include the combined results of ATLAS and CMS at 8~TeV~\cite{ATLAS:2016neq}, as well as the separate covariance matrices at 13~TeV, provided individually by ATLAS~\cite{ATLAS:2022vkf} and CMS~\cite{CMS:2022dwd}, since a combined analysis is not yet available. Additionally, we include the production and decay channels to charm-quark pairs, which were not accounted for in the aforementioned analyses~\cite{CMS:2022psv,ATLAS:2022ers}.
These observables put stringent constraints on the mixing angle and the alignment parameters, since the modifications of the Higgs couplings to weak gauge bosons and fermions in the A2HDM depend only on those parameters (see Eqs.~\eqref{eq:Higgs_weak_boson_couplings} and \eqref{eq:Higgs_yuk}).

\subsection{Direct detection}

Additional scalar particles beyond the SM (BSM) have been extensively searched for at different colliders. Due to the different collision energies, the LEP probed Higgsstrahlung-produced scalars up to masses of 120~GeV only, whereas the LHC has performed dedicated searches for both light and heavy scalars. To incorporate these results in our analysis, we compare the theoretically calculated production cross section of a given scalar particle, multiplied by its branching fraction to a particular decay channel, i.e. \sigbr, to the $95\%$ exclusion limit of the corresponding mode provided by the experimental collaborations. More specifically, we assign a normal distribution centred at $0$ (taking only positive values) to the ratio $\mathcal O_{\text{direct}}=\frac{(\sigma\cdot\mathcal B)_\text{A2HDM}}{(\sigma\cdot\mathcal B)_\text{exp}}$ between the theoretical estimate of \sigbr and its experimental upper bound. The variance of the density function of $\mathcal O_{\text{direct}}$ is adjusted in such a way that the value $1$ is disallowed at 95\% probability. We use a linear interpolation method to obtain a continuum from the discrete dataset given by the experiment. Concerning the theoretical predictions of the cross section for various processes, we have made extensive use of the open-source packages MadGraph5\_aMC@NLO~\cite{Alwall:2014hca}, HIGLU~\cite{Spira:1995mt} and HDECAY~\cite{Djouadi:2018xqq}, and the tabulated results available in CERN Yellow Reports~\cite{CERN8, CERN13}.

All the direct searches for BSM scalars included in our simulation are presented in Appendix \ref{sec:direct_searches}. 
For the light neutral scalars, we mostly take into account the ATLAS and CMS searches for a resonantly-produced 125~GeV Higgs decaying to two neutral scalars (see Tabs.~\ref{tab:CMS_light} and \ref{tab:ATLAS_light}). Apart from that, we incorporate associated production of light scalars with $t\bar t$, $b\bar b$, and weak gauge bosons, as given by ATLAS and CMS. The LHC results on a \cp-even neutral scalar decaying to two photons and a \cp-odd neutral scalar decaying to two taus are included too. 
In addition to the LHC data, we also consider the LEP upper bounds on the
pair production of two light neutral scalars and the associated production of one neutral scalar with a $Z$ boson through the Higgsstrahlung process (see Tab.~\ref{tab:LEP_neutral}).

Regarding the resonant production of a heavy scalar, we consider its decay to the different possible channels: $hh$, $hZ$, $Z$ boson associated with another heavy neutral scalar (see Tabs.~\ref{tab:CMS_heavy_scalar_to_scalar} and \ref{tab:ATLAS_heavy_scalar_to_scalar}), two gauge bosons (see Tabs.~\ref{tab:CMS_heavy_scalar_to_bosons} and \ref{tab:ATLAS_heavy_scalar_to_bosons}), and two fermions (see Tabs.~\ref{tab:CMS_heavy_scalar_to_fermions} and \ref{tab:ATLAS_heavy_scalar_to_fermions}), as presented by 
ATLAS and CMS. Secondary decays from these final states to SM particles are also mentioned in parentheses inside the tables. 

Concerning the searches for a light charged Higgs, we implemented the ATLAS and CMS results on a top quark decaying into the $H^+b$ mode (see Tabs.~\ref{tab:CMS_charged} and \ref{tab:ATLAS_charged}) and the LEP limits on the production of opposite-sign charged scalars through the $Z$-mediated s-channel (see Tab.~\ref{tab:LEP_charged}). The decay modes of the charged Higgs analysed in these channels are: $\tau^+\nu$, $c\bar s$, $c\bar b$, and $W^{+*} A$. On the other hand, for a charged scalar heavier than the top quark, we consider the results from ATLAS and CMS, pertaining to its resonant production followed by its decay to $\tau^+\nu$ and $t\bar b$ (see Tab.~\ref{tab:CMS_ATLAS_heavy_charged_scalar}).

The presence of light scalars may affect the decay widths, as well as the invisible widths, of several SM particles (like $W^\pm$, $Z$, the SM Higgs, and the top quark), which are well measured by experiments. Therefore, these observables put tight constraints on the model parameters. We have taken into account the current upper bounds on the invisible branching fraction of the Higgs boson and the invisible decay widths of the $W$ and $Z$ bosons, as well as the total decay width of the top quark, which are presented in Tab.~\ref{tab:widths}.

\subsection{Slepton searches}
\label{sec:slepton}

Many LEP and LHC analyses have been focused on the search for hypothetical SUSY particles. Since the decay of one slepton to a charged lepton and a neutralino resembles the collider signature of a charged Higgs decaying to a  charged lepton and a neutrino, one can try to adapt the slepton search data to the 2HDM context~\cite{Iguro:2023tbk}. The only difference is that neutrinos are almost massless while neutralinos could have arbitrary large masses.
Therefore, the LEP and LHC exclusion plots on slepton masses become relevant
for the charged Higgs search at very small neutralino masses.
In the absence of an official recasting by the experimental collaborations of their SUSY results into probes of generic scalars, in our analysis we take into account the  smuon and stau searches shown in Tab.~\ref{tab:SUSY_obs}, considering them as reinterpreted searches for the charged Higgs. It is, however, important to mention that only the searches for left-handed sleptons can be reinterpreted as searches for a charged Higgs because both of them have the same $SU(2)_L$ structure. The theoretical estimates for the pair-production cross section of left-handed sleptons (reinterpreted as $H^\pm$) through the Drell-Yan process at the LEP and the LHC (NLO+NLL) are taken from Refs.~\cite{Freitas:2003yp} and~\cite{Theo:susy, Fiaschi:2018xdm, Fuks:2013lya}, respectively.

\section{Fit setup}
\label{sec:Setup}

The numerical analysis in this work is performed using the open-source \texttt{HEPfit} package~\cite{DeBlas:2019ehy}, which employs a MCMC algorithm implemented via the \texttt{Bayesian Analysis Toolkit}~\cite{Caldwell:2008fw}. Renowned for its flexibility and computational efficiency, \texttt{HEPfit} has been widely utilised for global fits within both the SM~\cite{deBlas:2016ojx, deBlas:2021wap, deBlas:2022hdk} and various BSM frameworks, including effective field theories~\cite{DeBlas:2019qco, Durieux:2019rbz, Coutinho:2019aiy, Alasfar:2020mne, Crivellin:2020ebi, Husek:2020fru, Crivellin:2021njn, Darme:2021qzw, Miralles:2021dyw, Paul:2022dds,Miralles:2024huv} and specific new physics models~\cite{Cacchio:2016qyh, Chiang:2018cgb, Cheng:2018mkc, Eberhardt:2020dat, Eberhardt:2021ebh, Chen:2022zsh, Cheng:2022hbo, Chen:2023ins, Karan:2023kyj}. For this work, we have significantly modified and extended the existing implementation of the A2HDM in \HEPfit to enable the exploration of its low-mass regime. Information regarding the exact version of the code used in this analysis can be found in our statement of data availability, at the end of this paper.

Bayesian statistical inference is based on the interpretation of probability as an \textit{update of knowledge}, something that can be conjectured from the expression of Bayes' theorem: the so-called \textit{posterior} distribution of model parameters and observables follows from the statistics derived from all data, denoted as the \textit{likelihood}, times an initial probability density function of the full set of parameters, commonly termed \textit{priors}. By sampling from the likelihood distribution, multiplied by one's choice of \textit{a priori} probability distributions, which can be left flat if no particular initial knowledge --- either from experiments or theory computations --- exists for a given variable, one gets an updated set of values that reflects how new data has affected one's \textit{degree of belief}. For a comprehensive exposition of the inferential reasoning that informed the creation of UTfit and, subsequently, the development of \HEPfit, we refer to Refs.~\cite{DAgostini:1999gfj, Ciuchini:2000de}.

In a Bayesian framework, model comparison between different scenarios can be achieved by evaluating the Information Criterion (IC)~\cite{Ando:2007, Ando:2011}, defined as
\begin{equation}
    \text{IC} = -2\, \overline{\log \mathcal{L}} + 4\, \sigma^2_{\log \mathcal{L}} \, ,
\end{equation}
where $\overline{\log \mathcal{L}}$ is the average value of the \textit{log-likelihood}, and $\sigma^2_{\log \mathcal{L}}$ its variance. The IC is, thus, characterised by a first term which yields an estimate of the predictive accuracy of the model~\cite{Gelman:2013}, and a second term that serves as a penalty factor for the number of parameters used in the fit. The preference for a model against another is given according to which one has the smallest IC, with a suggested scale of evidence that can be read from Refs.~\cite{Jeffreys:1998, Kass:1995}.

\begin{table}[t!]
    \centering
\begin{tabular}{P{1.1cm}|P{1.1cm}|P{1.1cm}|P{1.1cm}|P{1.1cm}|P{1.1cm}|P{1.1cm}|P{1.1cm}|P{1.1cm}|P{1.1cm}|P{1.1cm}|P{1.1cm}|}
\hline
\multicolumn{12}{|c|}{\bf Priors} \\
\hline
\hline
\multicolumn{6}{|P{6.6 cm}}{$M_{\phi_{\text{light}}^{}} \in [10 \text{ GeV},\; M_h]$ } & \multicolumn{6}{|P{6.6 cm}|}{$M_{\phi_{\text{heavy}}^{}} \in [M_h,\; 700 \text{ GeV}]$ }  \\
\cline{2-12}
\multicolumn{4}{|P{4.4 cm}}{$\lambda_2\in [-1,\;10]$} & \multicolumn{4}{|P{4.4 cm}|}{$\lambda_3\in [-1,\;10]$}  & \multicolumn{4}{P{4.4 cm}|}{$\lambda_7\in [-3.5,\;3.5]$}  \\
\cline{2-12}
\multicolumn{3}{|P{3.3 cm}}{$\tilde \alpha\in [-0.2,\;0.2]$} & \multicolumn{3}{|P{3.3 cm}|}{$\varsigma_u\in [-0.5,\; 0.5]$} & \multicolumn{3}{P{3.3 cm}|}{$\varsigma_d \in [-10,\; 10]$} & \multicolumn{3}{P{3.3 cm}|}{$\varsigma_l \in [-100,\; 100]$} \\  
\hline
    \end{tabular}
    \caption{Priors chosen for the BSM parameters where $\phi_{\text{light}}^{}$ and $\phi_{\text{heavy}}^{}$ denote scalars lighter and heavier than the 125~GeV Higgs, respectively, with $\lbrace\phi_{\text{light}}^{},\; \phi_{\text{heavy}}^{}\rbrace\in \lbrace H,\; A,\; H^\pm\rbrace$.}
    \label{tab:prior}
\end{table}

\subsection{Selection of priors}
We now turn to the parameters we use in the fit, and our rationale for the priors that have been chosen for each. The \cp-conserving A2HDM introduces ten additional parameters beyond the SM: three scalar masses $(M_A, M_H$, and $M_{H^\pm})$, the mixing angle between the two \cp-even scalars $(\tilde\alpha)$, three independent quartic couplings ($\lambda_2, \lambda_3$, and $\lambda_7$), and three alignment parameters ($\varsigma_u, \varsigma_d$, and $\varsigma_l$). We have taken uniform distributions for the priors, and the ranges of them are chosen wide enough (mostly motivated by the stability and perturbativity bounds) to capture all the aspects of interesting physics signatures relevant for the study of light scalars. While choosing the ranges for the priors, shown in Tab.~\ref{tab:prior}, we have made sure that, after the global fits, there exist no significant statistics in the regions beyond the selected ranges, except for our requirement that forces at least one of the new scalars to be light.

There exists a great freedom regarding the choices of mass parameters for the BSM scalars. As already mentioned, we identify the physical state $h$ as the SM-like Higgs boson with mass $M_h=125.20$~GeV. The scenario with all the additional scalars heavier than the Higgs has been well studied in Refs.~\cite{Karan:2023kyj,Eberhardt:2020dat}. In this paper, we examine the possibility of the existence of scalars lighter than $M_h$. 
Seven distinct scenarios might appear, corresponding to different scalar mass ranges:
1) three cases with only one BSM scalar lighter than $M_h$,
2) three cases with two BSM scalars lighter than $M_h$, and
3) the case with all three BSM scalars lighter than $M_h$. 
We have performed global fits for the seven light-scalar scenarios, separately. As can be seen in Tab.~\ref{tab:prior}, we consider the mass of the light scalars to be within the range of 10~GeV to $M_h$, whereas the heavy scalar masses are taken to be between $M_h$ and 700~GeV. In principle, one could use as priors either the masses or their squares; however, as was already noticed in  Ref.~\cite{Eberhardt:2020dat}, the results obtained with the second choice turn out to be more sensitive to the chosen ranges of the priors. We thus restrict ourselves to the choice of linear priors only.

\section{Results}
\label{sec:Plots}

\renewcommand{\arraystretch}{1.7}

\begin{table}[ht!]
\begin{center}
\setlength{\arrayrulewidth}{0.8pt} 
\vspace*{-1.4cm}\hspace*{-1cm}\begin{tabular}{|c|P{1.1cm}|P{1.1cm}|P{1.1cm}|P{1.1cm}|P{1.1cm}|P{1.1cm}|P{1.1cm}|P{1.1cm}|P{1.1cm}|P{1.1cm}|P{1.1cm}|P{1.1cm}|}
\hline
\multicolumn{13}{|c|}{\bf Marginalised Individual Results} \\
\hline
\hline
\cellcolor{red!20!white}& \multicolumn{3}{|P{3.4 cm}}{IC:  84.06 } &\multicolumn{3}{|P{3.4 cm}}{$65\le M_{H} \le M_h$   } & \multicolumn{3}{|P{3.4 cm}|}{$168\le M_{A} \le 496$  } & \multicolumn{3}{P{3.4 cm}|}{$196\le M_{H^\pm} \le 500$  } \\
\cline{2-13}
\cellcolor{red!20!white}&\multicolumn{4}{|P{4.7 cm}}{$\lambda_2$ : $ 4.969 \pm 1.925$} & \multicolumn{4}{|P{4.8 cm}|}{$\lambda_3$ : $3.854\pm 2.067 $}  & \multicolumn{4}{P{4.7 cm}|}{$\lambda_7:$ $ 0.005 \pm 0.382$}  \\
\cline{2-13}
\cellcolor{red!20!white}\multirow{-3}{*}{\rotatebox{90}{$M_H\leq M_h$}}&\multicolumn{3}{|P{4.0 cm}}{$\tilde{\alpha}:$ $( 0.8\pm34.0 )\times10^{-3}$} & \multicolumn{3}{|P{3.3 cm}|}{$\varsigma_u:$ $ 0.001 \pm 0.073$} & \multicolumn{3}{P{3.3 cm}|}{$\varsigma_d:$ $0.017\pm 1.716 $} & \multicolumn{3}{P{3.3 cm}|}{$\varsigma_l:$ $ -0.325 \pm 20.100$} \\  
\hline
\hline
\cellcolor{blue!20!white}& \multicolumn{3}{|P{3.4 cm}}{IC:  83.74 } &\multicolumn{3}{|P{3.4 cm}}{$182\le M_{H} \le 500$  } & \multicolumn{3}{|P{3.4 cm}|}{$69\le M_{A} \le M_h$  } & \multicolumn{3}{P{3.4 cm}|}{$196\le M_{H^\pm} \le 500$  } \\
\cline{2-13}
\cellcolor{blue!20!white}&\multicolumn{4}{|P{4.4 cm}}{$\lambda_2$ : $4.609 \pm 1.891$} & \multicolumn{4}{|P{4.4 cm}|}{$\lambda_3$ : $3.817\pm 1.973$}  & \multicolumn{4}{P{4.4 cm}|}{$\lambda_7$ : $ 0.006\pm 1.164$}  \\
\cline{2-13}
\cellcolor{blue!20!white}\multirow{-3}{*}{\rotatebox{90}{$M_A\leq M_h$}}&\multicolumn{3}{|P{4.0 cm}}{$\tilde{\alpha}:$ $(-0.8 \pm 42.5)\times10^{-3}$} & \multicolumn{3}{|P{3.3 cm}|}{$\varsigma_u:$ $ 0.003\pm 0.106$} & \multicolumn{3}{P{3.3 cm}|}{$\varsigma_d:$ $ 0.008 \pm 1.348$} & \multicolumn{3}{P{3.3 cm}|}{$\varsigma_l:$ $ 0.105 \pm 15.380$} \\  
\hline 
\hline
\cellcolor{violet!30!white}& \multicolumn{3}{|P{3.4 cm}}{IC:  88.48 } &\multicolumn{3}{|P{3.4 cm}}{$M_h\le M_{H} \le 500$  } & \multicolumn{3}{|P{3.4 cm}|}{$M_h\le M_{A} \le 440$  } & \multicolumn{3}{P{3.4 cm}|}{$97\le M_{H^\pm} \le M_h$  } \\
\cline{2-13}
\cellcolor{violet!30!white}&\multicolumn{4}{|P{4.4 cm}}{$\lambda_2$ : $ 4.342\pm 2.185$} & \multicolumn{4}{|P{4.4 cm}|}{$\lambda_3$ : $ 0.280 \pm 0.214$}  & \multicolumn{4}{P{4.4 cm}|}{$\lambda_7:$ $0.004 \pm 0.873 $}  \\
\cline{2-13}
\cellcolor{violet!30!white}\multirow{-3}{*}{\rotatebox{90}{$M_{H^\pm}\leq M_h$}}&\multicolumn{3}{|P{4.0 cm}}{$\tilde{\alpha}:$ $( -1.7\pm 41.9)\times10^{-3}$} & \multicolumn{3}{|P{3.3 cm}|}{$\varsigma_u:$ $ 0.0006\pm 0.0364$} & \multicolumn{3}{P{3.3 cm}|}{$\varsigma_d:$ $ 0.004 \pm 0.706$} & \multicolumn{3}{P{3.3 cm}|}{$\varsigma_l:$ $ 0.528\pm 34.450$} \\  
\hline
\hline
\cellcolor{magenta!30!white}& \multicolumn{3}{|P{3.4 cm}}{IC:  89.48 } &\multicolumn{3}{|P{3.4 cm}}{$89\le M_{H} \le M_h$  } & \multicolumn{3}{|P{3.4 cm}|}{$78\le M_{A} \le M_h$  } & \multicolumn{3}{P{3.4 cm}|}{$154\le M_{H^\pm} \le 226$  } \\
\cline{2-13}
\cellcolor{magenta!30!white}&\multicolumn{4}{|P{4.4 cm}}{$\lambda_2$ : $ 4.890 \pm 2.166$} & \multicolumn{4}{|P{4.4 cm}|}{$\lambda_3$: $1.042\pm 0.5718 $}  & \multicolumn{4}{P{4.4 cm}|}{$\lambda_7:$ $0.002\pm 0.362 $}  \\
\cline{2-13}
\cellcolor{magenta!30!white}\multirow{-3}{*}{\rotatebox{90}{$M_{H,A}\leq M_h$}}&\multicolumn{3}{|P{4.0 cm}}{$\tilde{\alpha}:$ $(-0.6 \pm 47.4)\times10^{-3}$} & \multicolumn{3}{|P{3.3 cm}|}{$\varsigma_u:$ $ 0.002\pm 0.082$} & \multicolumn{3}{P{3.3 cm}|}{$\varsigma_d:$ $ 0.007\pm 1.620$} & \multicolumn{3}{P{3.3 cm}|}{$\varsigma_l:$ $ 0.370\pm 9.574$} \\  
\hline
\hline
\cellcolor{orange!30!white}& \multicolumn{3}{|P{3.4 cm}}{IC:  89.51 } &\multicolumn{3}{|P{3.4 cm}}{$85\le M_{H} \le M_h$  } & \multicolumn{3}{|P{3.4 cm}|}{$
M_h \le M_{A} \le 534$  } & \multicolumn{3}{P{3.4 cm}|}{$95\le M_{H^\pm} \le 120$  } \\
\cline{2-13}
\cellcolor{orange!30!white}&\multicolumn{4}{|P{4.4 cm}}{$\lambda_2$ : $ 4.639\pm 2.208$} & \multicolumn{4}{|P{4.4 cm}|}{$\lambda_3$ : $ 0.254\pm0.191$}  & \multicolumn{4}{P{4.4 cm}|}{$\lambda_7:$ $0.002\pm0.453 $}  \\
\cline{2-13}
\cellcolor{orange!30!white}\multirow{-3}{*}{\rotatebox{90}{$M_{H,H^\pm}\leq M_h$}}&\multicolumn{3}{|P{4.0 cm}}{$\tilde{\alpha}:$ $( 0.2\pm 38.2)\times10^{-3}$} & \multicolumn{3}{|P{3.6 cm}|}{$\varsigma_u:$ $ -0.0004\pm 0.0400$} & \multicolumn{3}{P{3.3 cm}|}{$\varsigma_d:$ $ -0.010\pm 0.656$} & \multicolumn{3}{P{3.3 cm}|}{$\varsigma_l:$ $ -0.603\pm 41.45$} \\  
\hline
\hline
\cellcolor{green!70!black!30!white}& \multicolumn{3}{|P{3.4 cm}}{\multirow{2}{*}{IC: 89.35}  } &\multicolumn{3}{|P{3.4 cm}}{$M_h\le M_{H} \le 163$} & \multicolumn{3}{|P{3.4 cm}|}{\multirow{2}{*}{$85\le M_{A} \le M_h$}} & \multicolumn{3}{P{3.4 cm}|}{\multirow{2}{*}{$95\le M_{H^\pm} \le 120$  }} \\\noalign{\vspace*{-3mm}}
\cellcolor{green!70!black!30!white}&\multicolumn{3}{|P{3.4 cm}}{   }&\multicolumn{3}{|P{3.4 cm}}{ $\cup\; 211\le M_{H} \le 553$ } & \multicolumn{3}{|P{3.4 cm}|}{  } & \multicolumn{3}{P{3.4 cm}|}{} \\
\cline{2-13}
\cellcolor{green!70!black!30!white}&\multicolumn{4}{|P{4.4 cm}}{$\lambda_2$ : $ 4.082\pm 2.157$} & \multicolumn{4}{|P{4.4 cm}|}{$\lambda_3$ : $ 0.246\pm0.192$}  & \multicolumn{4}{P{4.4 cm}|}{$\lambda_7:$ $-0.005\pm 0.993 $}  \\
\cline{2-13}
\cellcolor{green!70!black!30!white}\multirow{-3}{*}{\rotatebox{90}{$M_{A,H^\pm}\leq M_h$}}&\multicolumn{3}{|P{4.0 cm}}{$\tilde{\alpha}:$ $( -0.9\pm 41.0)\times10^{-3}$} & \multicolumn{3}{|P{3.6 cm}|}{$\varsigma_u:$ $ -0.0004\pm 0.0402$} & \multicolumn{3}{P{3.3 cm}|}{$\varsigma_d:$ $ 0.001\pm 0.779$} & \multicolumn{3}{P{3.3 cm}|}{$\varsigma_l:$ $ 0.261\pm 31.090$} \\  
\hline
\hline
\cellcolor{cyan!20!white}&\multicolumn{3}{|P{3.4 cm}}{IC: 89.22  }&\multicolumn{3}{|P{3.4 cm}}{$91\le M_{H} \le M_h$  } & \multicolumn{3}{|P{3.4 cm}|}{$83\le M_{A} \le M_h$  } & \multicolumn{3}{P{3.4 cm}|}{$95\le M_{H^\pm} \le 122$  } \\
\cline{2-13}
\cellcolor{cyan!20!white}&\multicolumn{4}{|P{4.4 cm}}{$\lambda_2$ : $ 4.740\pm 2.232$} & \multicolumn{4}{|P{4.4 cm}|}{$\lambda_3$ : $ 0.275 \pm 0.201$}  & \multicolumn{4}{P{4.4 cm}|}{$\lambda_7:$ $-0.002\pm 0.463 $}  \\
\cline{2-13}
\cellcolor{cyan!20!white}\multirow{-3}{*}{\rotatebox{90}{$M_{H,A,H^\pm}\leq M_h$}}&\multicolumn{3}{|P{4.0 cm}}{$\tilde{\alpha}:$ $( 0.4\pm38.3 )\times10^{-3}$} & \multicolumn{3}{|P{3.3 cm}|}{$\varsigma_u:$ $ -0.001\pm 0.043$} & \multicolumn{3}{P{3.3 cm}|}{$\varsigma_d:$ $ -0.005\pm 0.613$} & \multicolumn{3}{P{3.3 cm}|}{$\varsigma_l:$ $ -0.560\pm 41.890$} \\  
\hline
\end{tabular}
\caption{IC value and marginalised individual results. The mass limits are in GeV, at 95\% probability, while for the other parameters we show the mean value and the standard deviation.}
\label{tab:marginalised_results}
\end{center}
\end{table}

\renewcommand{\arraystretch}{1.4}

Here we show the results of the seven scenarios introduced in Sec.~\ref{sec:Setup}, which have been analysed in this work. After performing the fits, and despite imposing strong priors on the masses, we found that all the posterior probabilities of the observables, for all the scenarios, were compatible with the measurements within two standard deviations.\footnote{This would not be the case for $(g-2)_\mu$ when using the 2020 white paper prediction but, as explained before, we have not included this observable in our main result.} An immediate conclusion would be, then, that all scenarios contain a region of the parameter space compatible with all current data. Regarding the goodness of the fits, we found that all of them have similar values of the IC, as shown in Tab.~\ref{tab:marginalised_results}, meaning that the seven scenarios have a similar performance, though the cases with only one light neutral scalar perform slightly better. The IC values for the fits involving a single light neutral scalar are, in fact, remarkably close to those obtained from a global analysis that also allows all BSM scalars to be heavy (83.17), whose posteriors, in the end, match those from Ref.~\cite{Karan:2023kyj}.
The marginalised posterior distributions resulting from each fit are also summarised in Tab.~\ref{tab:marginalised_results}, with the information on masses presented in their 95\% probability ranges, while for the other model parameters we offer the mean value and the standard deviation. In general,
since at least one scalar was forced to be light, all scenarios prefer quite small values also for the mixing angle and the Yukawa alignment parameters with the absolute maxima at a 68\% probability: $\tilde{\alpha}\lesssim0.05$~rad, $\varsigma_u\lesssim0.1$, $\varsigma_u\lesssim1.7$, and $\varsigma_l\lesssim42$. However, the precise values for each scenario considered may differ significantly. In the following, we detail the correlations among the different parameters for all scenarios.

\begin{figure}[t!]
    \centering
    \hspace*{-0.4 cm}
    \includegraphics[scale=0.44]{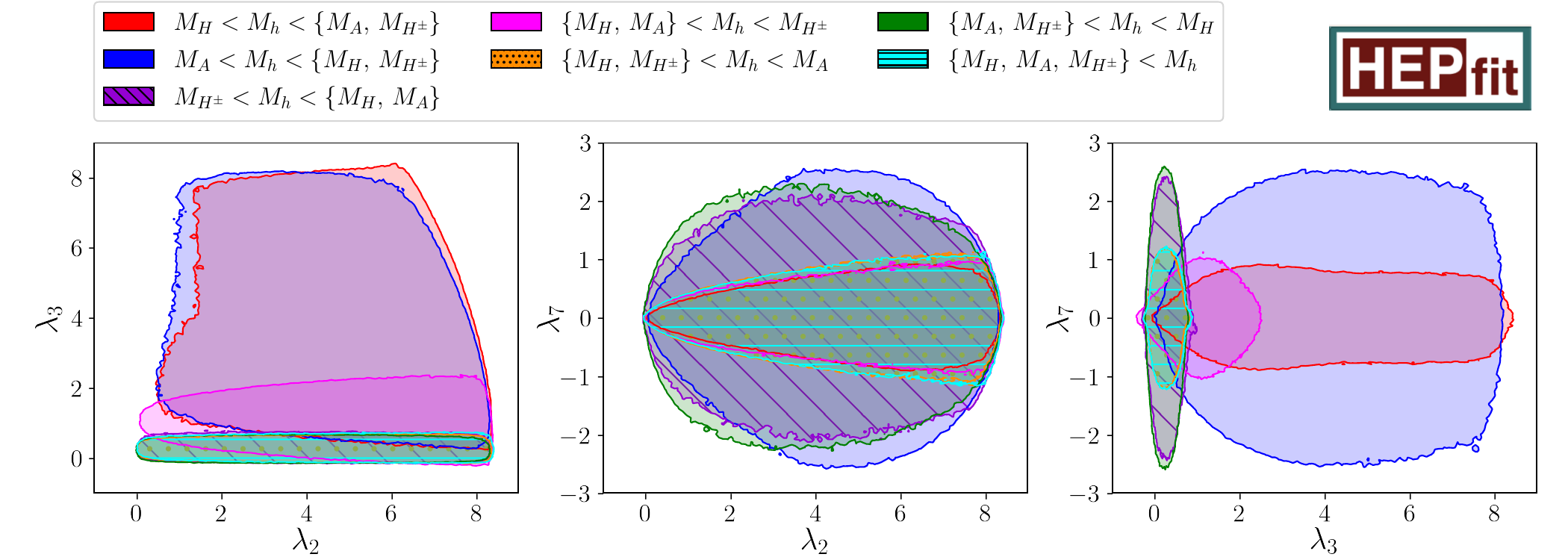}
    \caption{Correlations among the parameters of the scalar potential, shown as allowed regions with 95\% probability.}
    \label{fig:Lambdas}
\end{figure}

\subsection{Allowed regions for the model parameters}

\paragraph{Scalar potential couplings.-}
The correlations among the parameters of the potential are summarised in Fig.~\ref{fig:Lambdas} for all seven scenarios. The coupling $\lambda_2$ is only constrained by perturbative unitarity; therefore, its allowed regions are very similar in all scenarios, since they do not depend on the values of the masses. As shown by Eq.~\eqref{eq:pot_charge_neutral}, $\lambda_3$ and $\lambda_7$ are directly related to the couplings $\lambda_{h H^+ H^-}$ and
$\lambda_{H H^+ H^-}$; thus, they govern the decay of the \cp-even neutral scalars into $H^+ H^-$. Imposing the loop contributions to be smaller than the tree-level ones, as described in Sec.~\ref{sec:perturbativity}, provides, then, strong constraints on $\lambda_3$ and $\lambda_7$. These constraints become stronger when $M_{H^\pm} \ll M_{\varphi_i^0}$ , since the loop correction to $\lambda_{\varphi_i^0 H^+ H^-}$ scales as $1/M_{H^\pm}^2$, as can be seen in Eq.~\eqref{eq:pert_phiHpHp}. When $M_{H^\pm} \gg M_{\varphi_i^0}$, the loop correction scales as $1/M_{\varphi_i^0}^2$, and we also get a strong constraint on the coupling of that neutral scalar with the charged scalar.\footnote{The perturbative constraint is also relevant in this limit, since the effective coupling $(\lambda_{\varphi^0_iH^+H^-})_{\mathrm{eff}}$ determines the charged-scalar contribution to $\varphi_i^0\to\gamma\gamma$.} This explains the behaviour of the correlations in Fig.~\ref{fig:Lambdas}, where we see tight constraints on $\lambda_7$ when $M_H$ is small, while wider ranges are allowed in those scenarios with a heavier $M_H$ (blue, green, and purple regions). On the other hand, the constraints on $\lambda_3$ are looser when the charged scalar can be heavy (red, blue, and magenta regions). Note that in the magenta region, although we allow $M_{H^\pm}$ to be heavy, the oblique parameters force the charged scalar to also be light because the other two scalars are both light.

\begin{figure}[t!]
    \centering
    \hspace*{-0.4 cm}
    \includegraphics[scale=0.44]{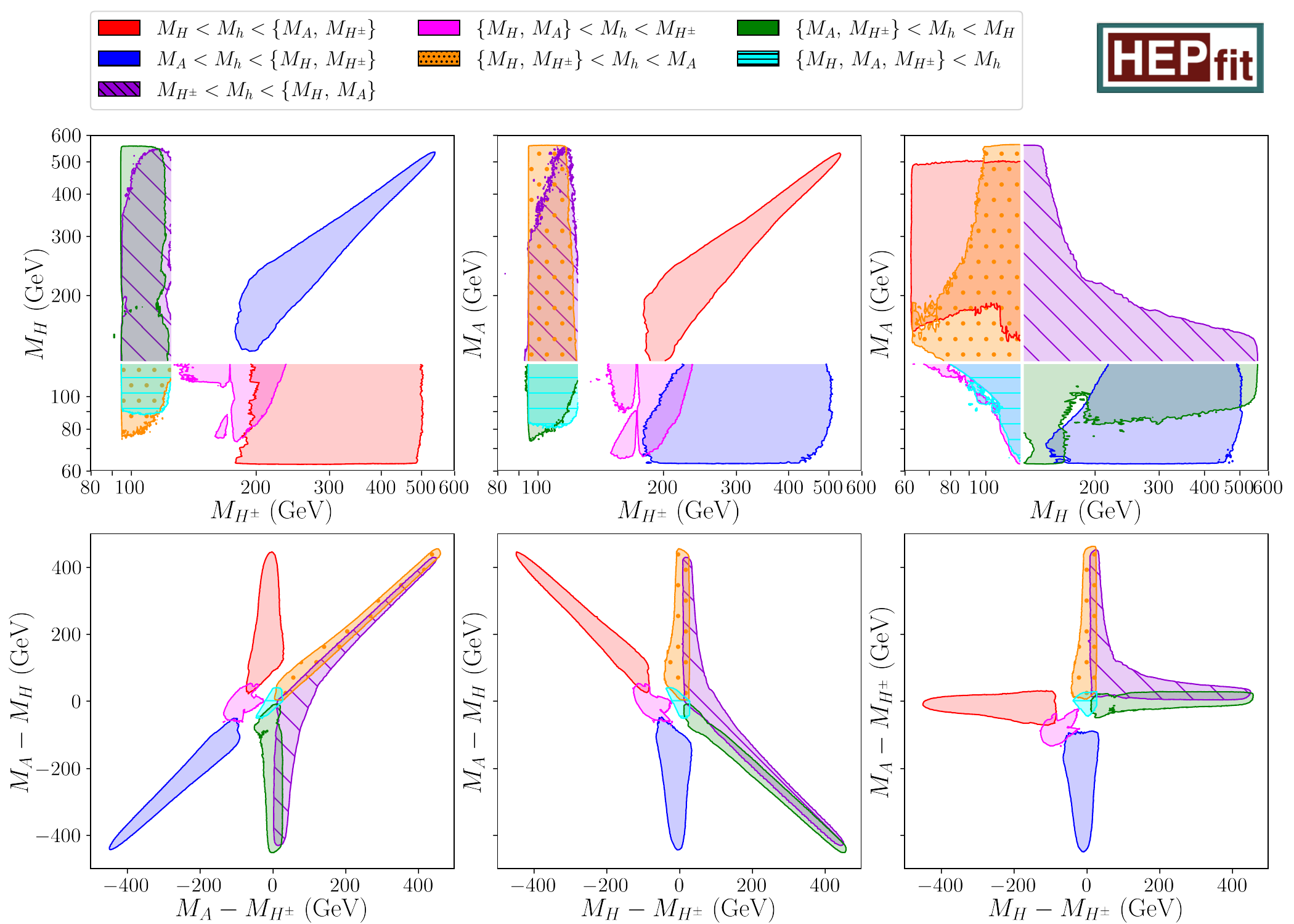}
    \caption{Correlations among the scalar masses and their mass splitting, shown as 95\% allowed probability regions.}
    \label{fig:masses}
\end{figure}

\paragraph{Scalar masses.-} 
Fig.~\ref{fig:masses} shows the correlations among the scalar masses, and the correlations among their mass splitting. There is a clear complementarity between the different scenarios considered, which is particularly noticeable in the correlation among the two neutral scalars. In order to satisfy the constraints from the oblique parameters, the mass of the charged scalar must align with one of the neutral scalar masses, hence the complementarity. When the two neutral scalars are forced to be light (magenta), $M_{H^\pm}$ is also pushed to be small, even though we allow it to be heavy, as mentioned before. If we impose one of the neutral scalars to be light and the other to be heavy together with the charged scalar (blue/red), the splitting between the two heavy masses becomes small. When only the charged scalar is imposed to be light (purple), one of the two neutral scalars must also be light to align with the charged one. However, when the charged scalar is imposed to be light together with a neutral scalar (orange/green), the mass of the other neutral scalar can reach large values, as in those cases the oblique constraints are avoided with the alignment of the masses of the two light scalars. 

In the scenario with three light scalars (cyan), the splitting of their masses is necessarily small, and for this reason the allowed region is found to be confined to the centre of all correlation plots among the scalar mass differences. Similarly, the scenario with two light neutral scalars (magenta) is also found to be in the central region of the same plots. Concerning the other scenarios, the alignment of the allowed regions along the mass-splitting axes can easily be understood from the corresponding ranges of allowed masses.

\begin{figure}[ht!]
    \centering
    \hspace*{-0.8 cm}
    \includegraphics[scale=0.44]{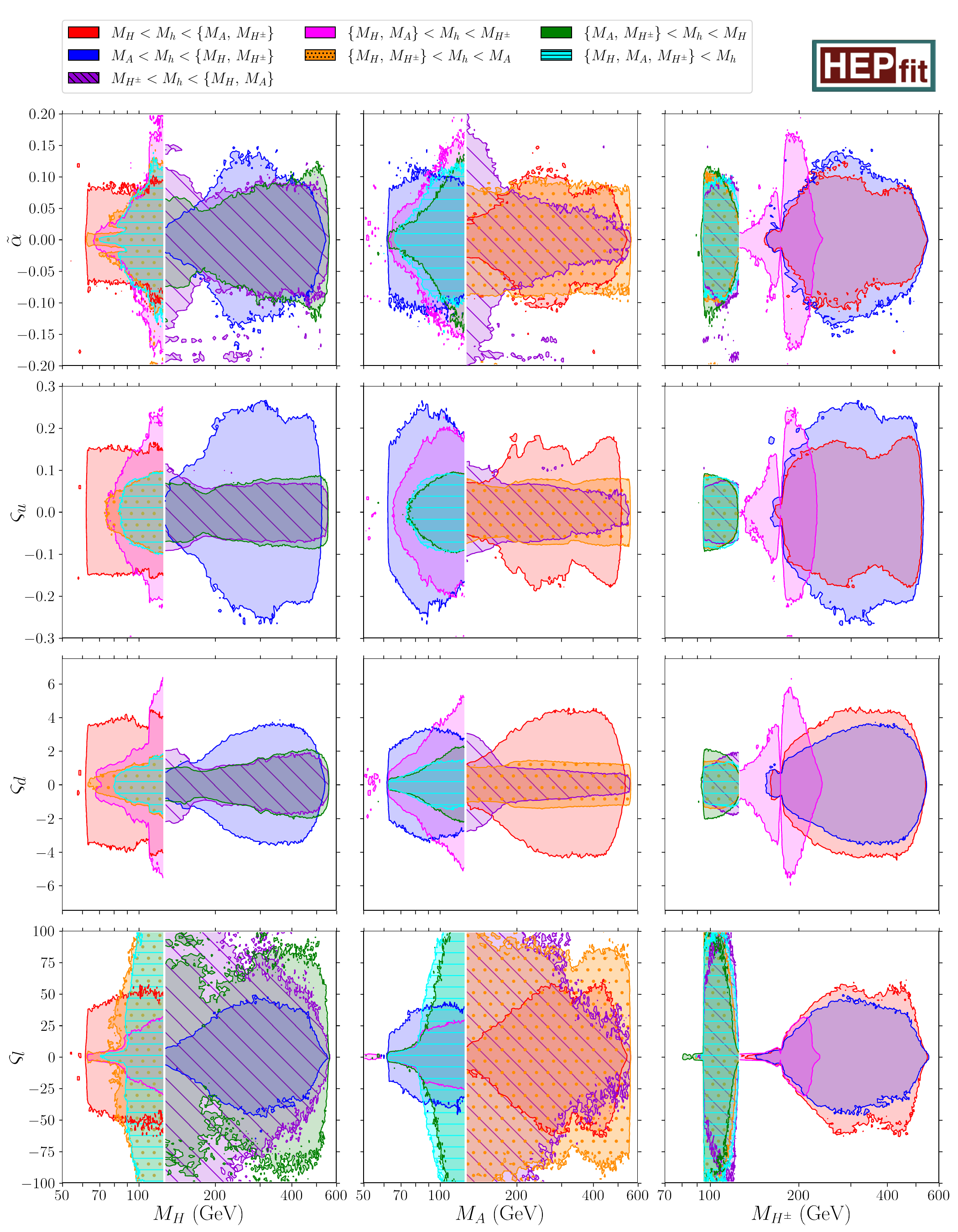}
    \caption{Correlations between the mixing angle (in rad) and the Yukawa alignment parameters with the scalar masses, shown as 95\% allowed probability regions.}
    \label{fig:Masses_vs_alpha_and_couplings}
\end{figure}

The theoretical constraints limit the maximum mass difference among all the scalars, disfavouring masses above 600~GeV when any scalar is lighter than the Higgs boson. The direct searches also play their role in imposing tight bounds on the masses, forcing the charged scalar to always be heavier than 90~GeV and the neutral scalars to be heavier than 60~GeV. These features are clearly seen in the correlation among the masses and the Yukawa alignment parameters, shown in Fig.~\ref{fig:Masses_vs_alpha_and_couplings}.

\paragraph{Masses versus mixing angle and Yukawa alignment parameters.-} As illustrated in Fig.~\ref{fig:Masses_vs_alpha_and_couplings}, the constraint that neutral masses are to be heavier than 60~GeV cannot be avoided, even for small values of the other parameters. This limit arises because masses below 60~GeV would enable the 125~GeV Higgs to decay into neutral scalars, a decay channel that has been extensively scrutinised by LHC experiments. By the same token, charged scalars lighter than 95~GeV are tightly constrained by the LEP searches for double production of charged scalars decaying to taus and quarks. Remarkably, for masses between 95~GeV and around 120~GeV 
the constraints on the leptonic coupling from direct searches exhibit a gap,
which manifests as an allowed region including extremely high values for $\varsigma_l$. This region might have been scrutinised by the slepton searches that we have added in this analysis but, unfortunately, the slepton limits start above 120~GeV, and this is precisely why below that value we see the wider allowed region. In principle, flavour observables could also constrain this region, but the current limits on $\varsigma_l$ are very weak.
In those scenarios with a light charged scalar, flavour provides strong constraints on $\varsigma_u$ and $\varsigma_d$, pushing these parameters to be small; this reduces significantly the sensitivity to the leptonic alignment parameter, so that values of  $\varsigma_l$ as high as 100 become allowed.

\begin{figure}[t!]
    \centering
    \hspace*{-0.5 cm}
    \includegraphics[scale=0.44]{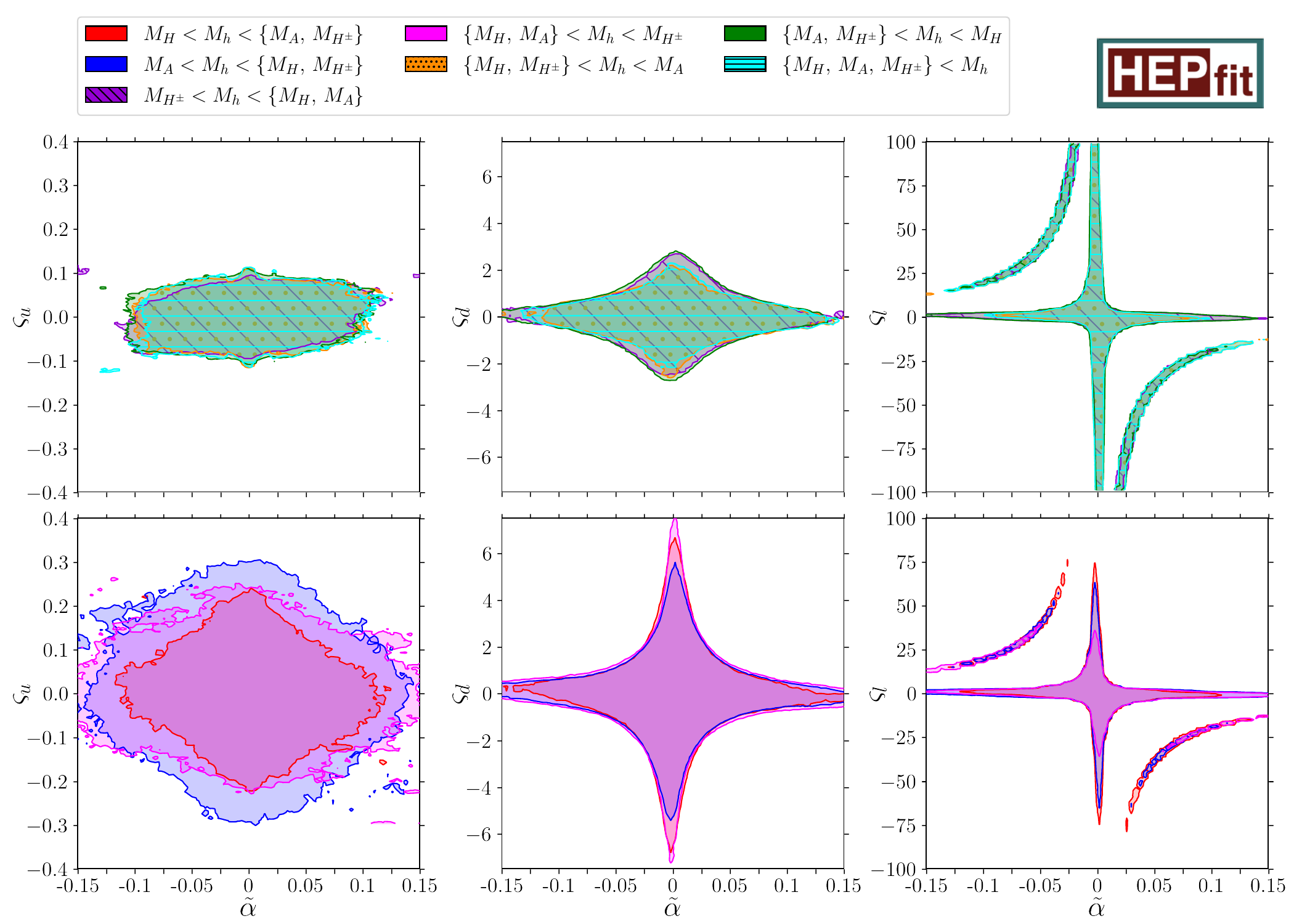}
    \caption{Correlations among the mixing angle (in rad) and the Yukawa alignment parameters, shown as 95\% allowed probability regions. The panels are duplicated to facilitate readability. In the top panel we present the scenarios with $M_{H^\pm}$ light, the rest are presented in the bottom panel.}
    \label{fig:Couplings_vs_alpha}
\end{figure}
 
The dominant BSM contributions to flavour observables are mostly proportional to $\varsigma_u^2 m_t^2/M_{H^\pm}^2$, $\varsigma_d^2 m_b^2/M_{H^\pm}^2$, and $\varsigma_u\varsigma_d m_t m_b/M_{H^\pm}^2$,
which enforces tight constraints on $\varsigma_u$ and somewhat softer upper bounds on $\varsigma_d$. The allowed ranges for these two parameters get obviously broader at larger values of $M_{H^{\pm}}$ (red, blue, and magenta regions). On the other hand, the LHC direct-detection observables restrict the combinations $\varsigma_u \varsigma_l$ and $\varsigma_d \varsigma_l$, where the $\varsigma_{u,d}$ parts come from the production channels involving quark-scalar vertices, and the $\varsigma_l$ part arises from the leptonic decay of the scalars. Therefore, the scenarios with larger masses of the charged scalar indirectly restrict $\varsigma_l$ to smaller ranges (red, blue, and magenta regions), whereas the scenarios with lighter $M_{H^\pm}$ allow for comparatively larger values of $\varsigma_l$ (purple, orange, green, and cyan regions).

Owing to the good agreement of the Higgs signal-strengths with the SM expectations, the mixing $\tilde\alpha$ between the two \cp-even neutral scalars is bounded to be small in all scenarios, a visual confirmation, at the 95\% probability region, of what we have already reported at the start of this section concerning the 68\% probability ranges we obtained for this parameter.

\begin{figure}[t!]
    \centering
    \hspace*{-0.5 cm}
    \includegraphics[scale=0.44]{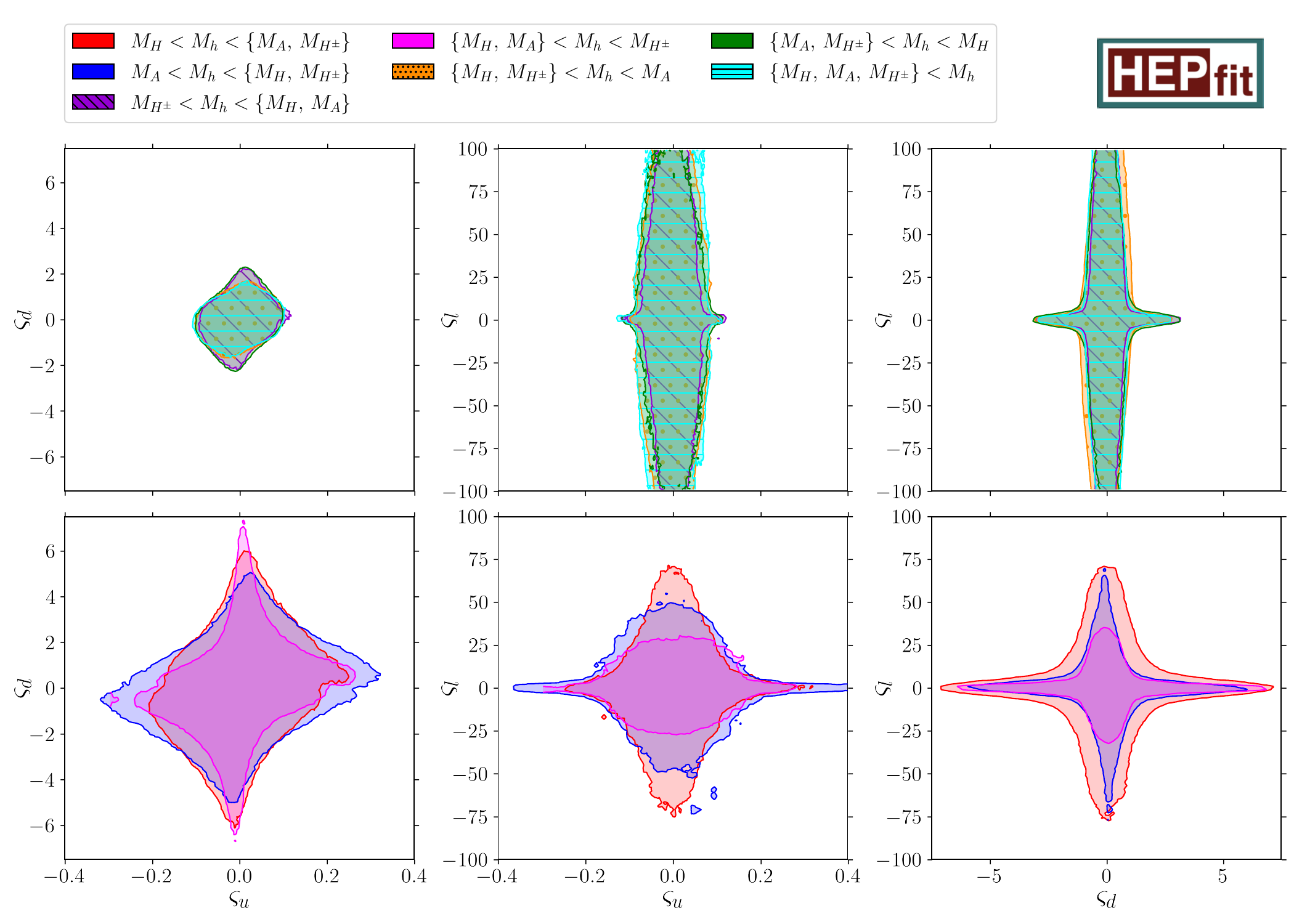}
    \caption{Correlations among the Yukawa alignment parameters, shown as 95\% allowed probability regions. To facilitate readability, the results are provided with the same split in $M_{H^\pm}$ scenarios as in Fig.~\ref{fig:Couplings_vs_alpha}.}
    \label{fig:Couplings}
\end{figure}

\paragraph{Mixing angle versus Yukawa alignment parameters.-} As shown in Fig.~\ref{fig:Couplings_vs_alpha}, the mixing angle is significantly correlated with the Yukawa alignment parameters. While the measured Higgs coupling to the gauge bosons puts an absolute lower bound on $\cos{\tilde\alpha}$, the Yukawa couplings in Eq.~\eqref{eq:Higgs_yuk} are sensitive to $\varsigma_f \sin{\tilde\alpha}$, providing a stronger constraint on $\tilde\alpha$ unless the alignment parameters are very small. 
Once these parameters are set to small values, the Higgs Yukawa couplings 
become extremely close to the SM prediction, for the values of the mixing angle considered here. The correlation with the up-type alignment parameter, shown in the left panels of Fig.~\ref{fig:Couplings_vs_alpha}, is broad, since the flavour observables already impose strong constraints on $\varsigma_u$, suppressing its contribution to the signal strengths. 
The correlations are more visible in the centre ($\varsigma_d$) and right ($\varsigma_l$) panels. The leptonic alignment parameter exhibits two additional solutions where the measured absolute value of the Yukawa coupling is reached through a destructive interference between the $\cos{\tilde\alpha}$ and $\varsigma_l\sin{\tilde\alpha}$ contributions. This flipped-sign solution is not viable for the quark Yukawas, which are subject to strong constraints from other observables (like flavour and direct searches). As already pointed out, the scenarios with heavier ${H^\pm}$ (red, blue, and magenta) are found to have larger allowed ranges of $\varsigma_{u,d}$ and smaller ones of $\varsigma_l$.

\paragraph{Yukawa alignment parameters.-} The correlations of the Yukawa alignment parameters among themselves are shown in Fig.~\ref{fig:Couplings}. We can clearly see that, in the scenarios with a light charged scalar, both $\varsigma_u$ and $\varsigma_d$ are forced to be small, while much looser constraints are found for $\varsigma_l$. As rationalised before, with such small values of the quark alignment parameters the flavour constraints become less sensitive to the lepton coupling. The correlations of the leptonic coupling with the quark ones clearly indicate that:
in order to obtain values of $\varsigma_l$ as big as 100, the model needs to have very small values of $\varsigma_u\lesssim 0.1$ and $\varsigma_d\lesssim 1$; and in the scenarios where the quark couplings are allowed a bit more of free rein, $\varsigma_l$ compensates by getting increasingly squished in those regions of parameter space.

\subsection{Effect of slepton searches}
\label{sec:susy}

As already mentioned in Sec. \ref{sec:slepton}, (left-handed) slepton searches at colliders with small neutralino masses can be reinterpreted as the search for a charged scalar decaying to a lepton and a neutrino
because the two processes exhibit the same dilepton plus missing-energy signature in the final state. Although in our global fits we have incorporated the LEP and LHC slepton searches, in this subsection we address the influence they have on our results. There is a striking phenomenological difference between slepton and charged Higgs searches at colliders: the slepton decay usually conserves the lepton flavour, with a maximum of $5\% - 10\%$ branching fraction permitted to lepton-flavour-violating decays~\cite{Bartl:2005yy} --- as such, the experimental searches for sleptons assume lepton flavour conservation, i.e., a 100\% probability to decay into a particular lepton. Yet, apart from a small branching fraction to $c\bar b$, a light charged Higgs has two dominant decay modes: $\tau^+\nu$ and $c \bar s$; if enough phase space is available,
it can also decay significantly to $t\bar b$, and even $W^{+*}A$. After taking these branching fractions into account, the total cross section for the production of a pair of charged scalars and their subsequent decay into two specific leptons with missing energy may become substantially smaller than the one for the analogous slepton signature. Since the coupling of a charged Higgs to a charged lepton and a neutrino is proportional to the mass of the charged lepton in play, its electronic and muonic branching fractions 
are negligible and, thus, the selectron and smuon searches from LEP and ATLAS (referenced in Tab.~\ref{tab:SUSY_obs}) put quite weak bounds on the parameters involved in charged Higgs phenomenology. The LEP bounds from stau searches are also weaker than the constraints arising from the direct LEP searches for a charged Higgs decaying to $\tau^+\nu$. Only the LHC stau searches have some non-negligible effect on the charged scalar parameters. However, the more stringent constraint on a light charged Higgs comes from the CMS $\tilde \tau_L$ search, which only applies for
charged scalar masses heavier than 115~GeV~\cite{CMS:2022syk}.

\begin{figure}[t!]
    \centering
    \includegraphics[scale=0.44]{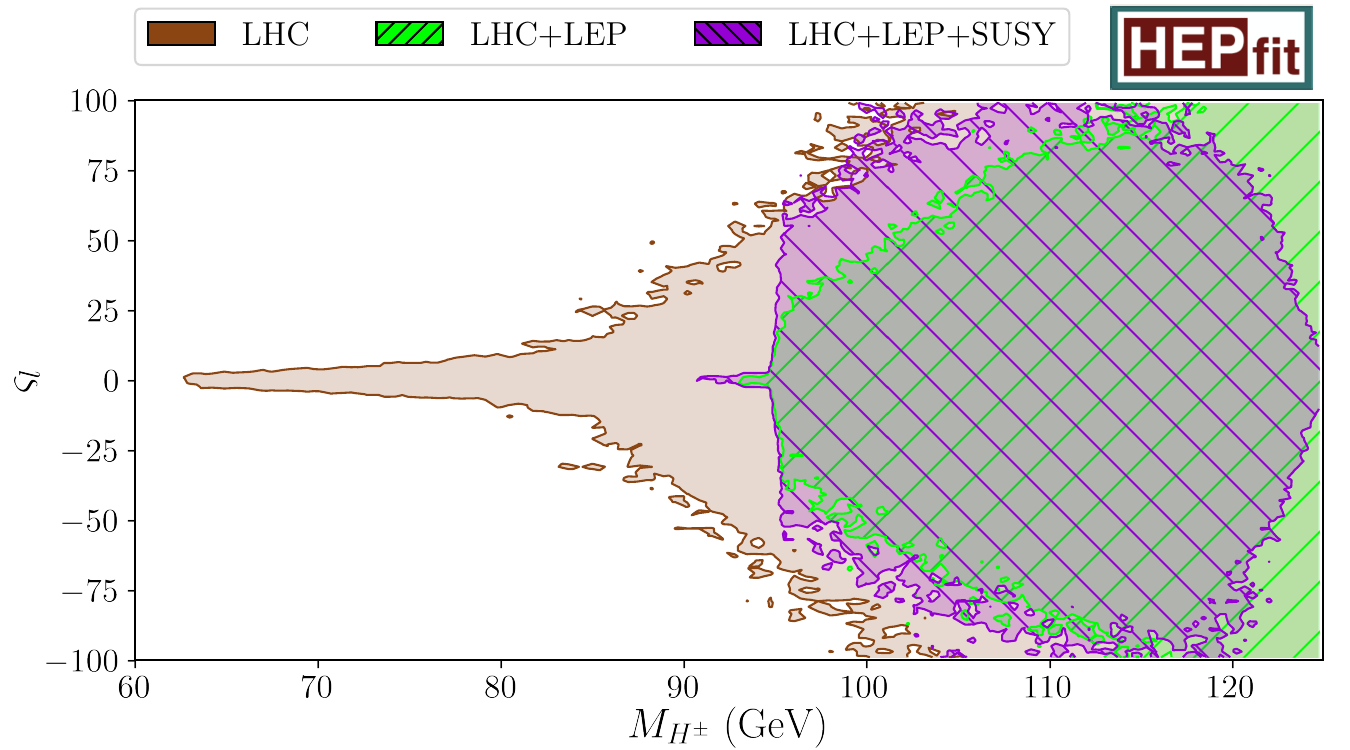}
    \caption{Correlation among $\varsigma_l$ and $M_{H^\pm}$ after successive applications of direct search constraints.
    The region with all collider data included (purple) matches that from the global fit, shown in a different aspect ratio in Fig.~\ref{fig:Masses_vs_alpha_and_couplings}.}
    \label{fig:mHp_cmparison}
\end{figure}

Fig.~\ref{fig:mHp_cmparison} compares the constraints on the $\varsigma_l-M_{H^\pm}$ correlation emerging from direct searches at the LHC (all other constraints apart from the direct searches are also included) with those obtained after adding successively the LEP scalar searches and the SUSY searches. As can be seen, a global fit with direct searches from the LHC only (brown region) restricts the lower bound on the charged Higgs mass to 62.5~GeV (i.e. $M_h/2$), whereas the fit with direct search data from the LEP and the LHC (green region) pushes the lower bound on $M_{H^\pm}$ to $\sim\! 95$~GeV.  
The slepton searches (purple) remove  regions with higher values of $|\varsigma_l|$ for $115~\text{GeV}<M_{H^\pm}<M_h$. As a result of constraining the masses above $115~\text{GeV}$, the minimum of the likelihood shifts toward smaller $ M_{H^\pm} $, where limits sensitive to $|\varsigma_l|$ from direct searches are absent. A consequence of this shift is that, at 95\% probability, slightly higher values of $|\varsigma_l|$ become allowed when including the SUSY searches. Nevertheless, this effect diminishes on the regions with higher probability.

Because of the CMS $\tilde \tau_L$ search, the marginalised upper bound at 95\% probability on a light charged Higgs mass does not reach $M_h$ in some cases, as can be noticed from the last three scenarios in Tab.~\ref{tab:marginalised_results}.

\section{\boldmath Implications of $(g-2)_\mu$}
\label{sec:gm2}

Besides the set of flavour observables that was used to produce the results presented in the previous section, \HEPfit contains a full implementation of the NP contribution to the anomalous magnetic moment of the muon, $a_\mu=(g-2)_\mu/2$, based on Ref.~\cite{Ilisie:2015tra}, including all relevant 1- and 2-loop contributions for the A2HDM~\cite{Lautrup:1971jf, Leveille:1977rc, Czarnecki:1995wq, Dedes:2001nx, Chang:2000ii, Cheung:2001hz, Cheung:2003pw,  Cherchiglia:2016eui, Athron:2021evk}.
Unfortunately, the SM prediction of $a_\mu$ remains very uncertain due to a significant discrepancy between the estimated hadronic vacuum polarisation contribution derived from data-driven approaches, using dispersive $e^+e^-$ data~\cite{Aoyama:2020ynm}, and those obtained from $\tau$-decay data~\cite{Davier:2023fpl,Masjuan:2023qsp} or lattice QCD calculations, which can be seen in collected form in the publication with the latest computation from the BMW group~\cite{Boccaletti:2024guq}. The recent  CMD-3 high-precision measurement of the $e^+e^-\to\pi^+\pi^-$ cross section~\cite{CMD-3:2023alj, CMD-3:2023rfe} shows in fact a very good compatibility with $\tau$ and lattice data, implying a much better agreement with the muon $g-2$ measurement than previously estimated from $e^+e^-$ data. Nevertheless, given all these discrepancies and the lack of consensus within the community, we have chosen not to include this observable in the main fit. However, we have evaluated its impact separately. 
In order to illustrate the implications of the different theoretical predictions, we have performed two different global fits: 
\begin{enumerate}
    \item Keeping all theoretical estimates from the 2020 white paper unchanged~\cite{Aoyama:2020ynm}, but updating the experimental value of $(g-2)_\mu$ to the most recent world average~\cite{Muong-2:2023cdq, Muong-2:2024hpx}, such that the difference between both becomes
    \begin{equation} 
    \Delta a_\mu^{\text{WP}} = a_\mu^{\text{exp}} - a_\mu^{\text{SM, WP}} = 249\; (49) \times 10^{-11} \, .
    \label{eq:gm2_inputs_WP}
    \end{equation}
    This claimed $5\sigma$ \qt{anomaly} has triggered a large number of NP explanations, some of them involving light scalars~\cite{Wang:2014sda, Ilisie:2015tra, Abe:2015oca, Han:2015yys, Cherchiglia:2017uwv, Athron:2021iuf, Jueid:2021avn, Dey:2021pyn, Botella:2022rte, Iguro:2023tbk, Afik:2023vyl,Doff:2024hid}.
    \item Adopting the latest BMW computations of the hadronic vacuum polarisation~\cite{Boccaletti:2024guq} and light-by-light scattering~\cite{Fodor:2024jyn} contributions,
    \begin{equation} 
    \Delta a_\mu^{\text{BMW}} = a_\mu^{\text{exp}} - a_\mu^{\text{SM, BMW}} = 4\; (42) \times 10^{-11} \, ,
    \label{eq:gm2_inputs_BMW}
    \end{equation}
    in excellent agreement with the experimental measurement.
\end{enumerate}

\begin{figure}[t!]
    \centering
    \includegraphics[scale=0.44]{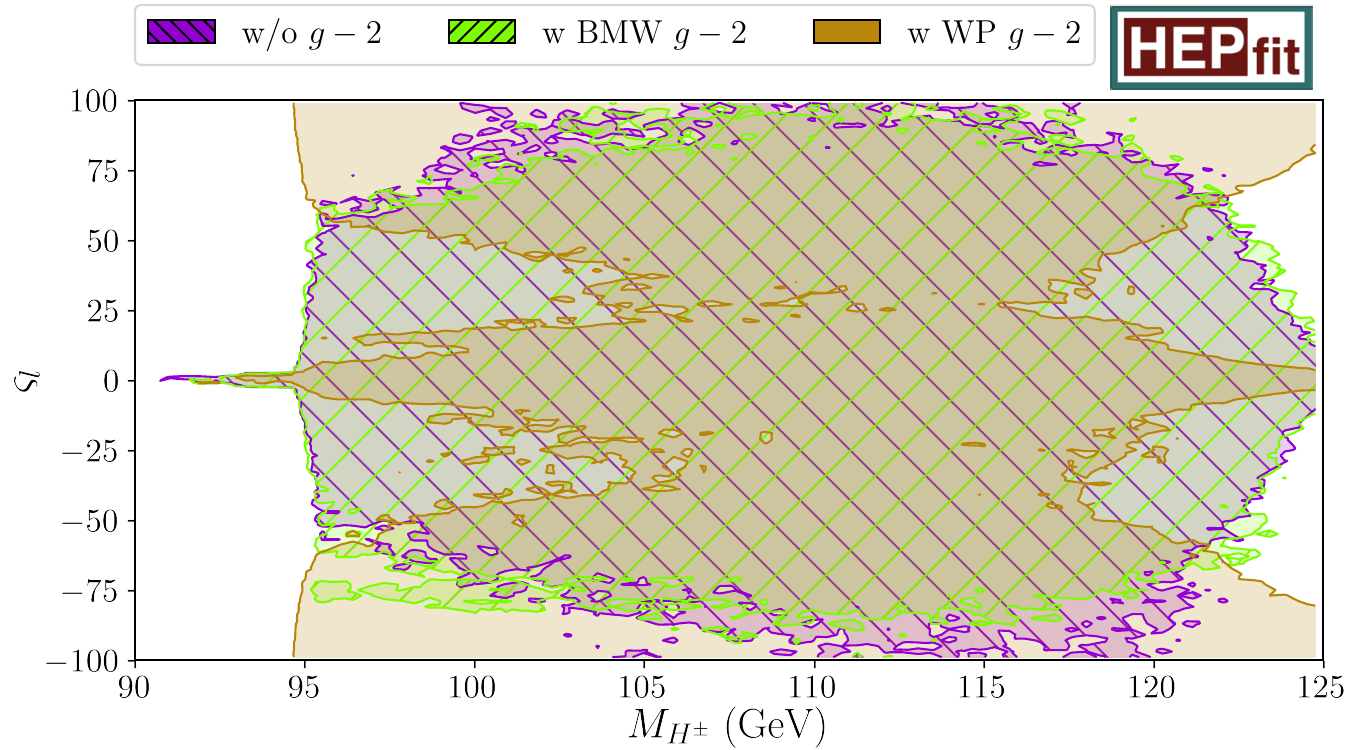}
    \caption{ Allowed 95\% probability regions for $\varsigma_l$ versus $M_{H^\pm}$ for the global fit with the prior 
    $M_{{H^\pm}}\le M_h\le M_{H,A}$, without including the observable $(g-2)_\mu$ (purple, as in Fig.~\ref{fig:Masses_vs_alpha_and_couplings}), adding the prediction of $(g-2)_\mu$ from the 2020 white paper (brown), and adding the prediction of $(g-2)_\mu$ from the BMW collaboration (green). The inputs used for the white paper and BMW scenarios can be found in Eqs.~\eqref{eq:gm2_inputs_WP} and \eqref{eq:gm2_inputs_BMW}, respectively.}
    \label{fig:gm2_comp}
\end{figure}

Following this initial exposition, we now assess how each $\Delta a_\mu$ above would affect our global analysis. 2HDM explanations of the so-called \qt{$g-2$ anomaly} require light scalar or pseudoscalar particles with large values of their leptonic couplings~\cite{Ilisie:2015tra}.
In order to maximise the possible effect, we focus on scenarios with $M_{H^\pm} < M_h$ because in this case there are insufficient observables to constrain $\varsigma_l$ below the perturbative limit of 100, at a 95\% probability. Fig.~\ref{fig:gm2_comp} compares in the $M_{H^\pm}-\varsigma_l$ plane the allowed 95\% probability region obtained without including $(g-2)_\mu$ (purple) with the ones emerging from global fits including $(g-2)_\mu$: either with the 2020 white paper (brown) or the latest BMW (green) SM predictions.

The impact of $(g-2)_\mu$ in the global fit is negligible when the SM theoretical prediction agrees with the experimental measurement because, with the current uncertainty, this observable is not able to further constrain $\varsigma_l$. Since the BMW value is completely compatible with the experimental measurement, there is a perfect overlap between the purple and green regions in the figure. 

The situation is very different if one employs the white paper prediction, which deviates from the experimental value by around 5 standard deviations.
Reducing the tension with $(g-2)_\mu$ requires then larger values of $\varsigma_l$. However, since all the other observables prefer smaller values of this parameter, a region with very small values of $\varsigma_l$ remains allowed at 95\% probability. 
For this last region, the prediction for $(g-2)_\mu$ is obviously in huge tension, but the tension for all the other observables is reduced, while the opposite happens for the region with larger values of $\varsigma_l$.
This tension translates into a much worse performance of the fit.
Indeed, the IC value for this fit is around 129, while for the fit with BMW data and for the fit without $(g-2)_\mu$, it is about 88. As evident from this difference of $\sim\! 40$ in ICs, which reflects the comparison of hypotheses as would be obtained with the calculation of the Bayes factor, the A2HDM model is clearly not a great solution to explain the tension between the $(g-2)_\mu$ predicted by the white paper and the experimental measurement.

\begin{figure}[t!]
    \centering
    \includegraphics[scale=0.44]{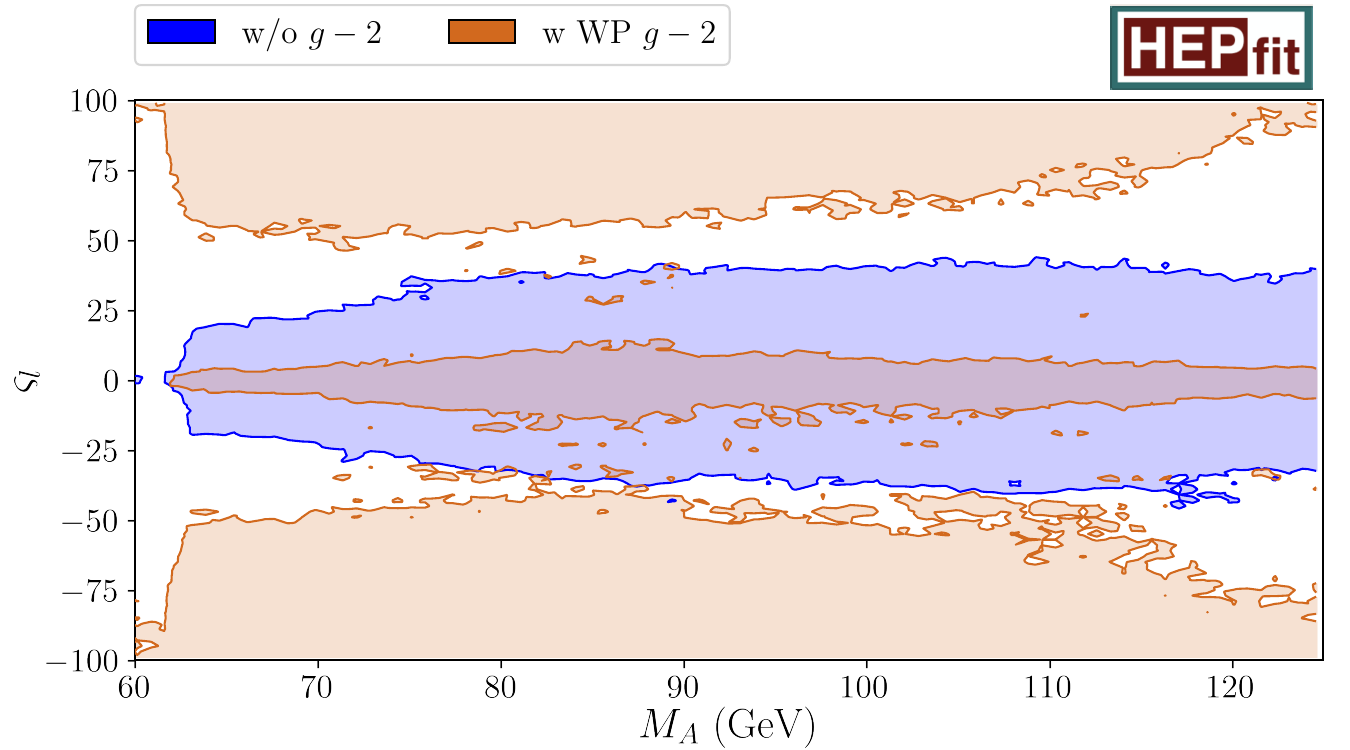}
    \caption{ Allowed 95\% probability regions for $\varsigma_l$ versus $M_{A}$ for the global fit with the prior 
    $M_{{A}}\le M_h\le M_{H,H^\pm}$, without including the observable $(g-2)_\mu$ (blue, as in Fig.~\ref{fig:Masses_vs_alpha_and_couplings}) and adding the prediction of $(g-2)_\mu$ from the 2020 white paper (brown). The inputs used for the white paper scenario can be found in
    Eq.~\eqref{eq:gm2_inputs_WP}. }
    \label{fig:gm2_comp_A}
\end{figure}

The situation is similar for the other scenarios. Incorporating $(g-2)_\mu$ using the theoretical prediction of the white paper deteriorates significantly the quality of the fits. To provide another example, we compare, in Fig.~\ref{fig:gm2_comp_A}, the corresponding fits for the case of a light pseudoscalar, a widely used model to explain the $(g-2)_\mu$ anomaly (see, e.g., the review presented in Ref.~\cite{Iguro:2023tbk}, and references therein). As found in the case of a light charged scalar, two clearly separated regions appear in this case, in such a way that performing an analysis with the input from the white paper generates also here a strong tension with the data. Therefore, a possible explanation of this anomaly with a light pseudoscalar within the A2HDM is, as is the case for any extra scalar from this model, extremely disfavoured once enough data are taken into account.

\section{Conclusions}
\label{sec:Concl}

In this paper we have studied in great detail whether neutral and/or charged scalars lighter than the 125~GeV Higgs could be compatible with current data, within the framework of the A2HDM. The \cp-conserving A2HDM introduces ten independent BSM parameters: the masses of three new scalars ($M_H, M_A$, and $M_{H^\pm}$), three scalar-potential parameters ($\lambda_2, \lambda_3$, and $\lambda_7$), three alignment parameters ($\varsigma_u, \varsigma_d$, and $\varsigma_l$), and the angle $\tilde\alpha$ mixing the two \cp-even scalars. The following constraints have been imposed to restrict the parameter space of the model in the light-mass regime: vacuum stability, perturbativity, electroweak precision observables, flavour observables, Higgs signal strengths, and several relevant direct searches at high-energy colliders (LEP and LHC).
We have performed our fits with the help of the open-source code \HEPfit, which employs Bayesian statistics powered by a MCMC algorithm. Depending on the ranges attributed to the masses of the BSM scalars, seven different scenarios arise. According to the IC values obtained for each fit, all these seven scenarios show a similar adeptness at fitting the data, though the scenarios with only one light neutral scalar perform marginally better --- in fact, the IC values of the latter are even identical to those from a fit with all scalars allowed to be heavy.

The measured (125~GeV) Higgs-boson properties at the LHC exclude that the neutral scalars could be lighter than half of the Higgs mass, at 95\% probability. On the other hand, the LEP searches push the lower bound of the charged scalar mass to 95~GeV at the same probability. In all these light scenarios, the bounds from stability, perturbativity, and electroweak precision observables never allow the masses of the heavier scalars to reach beyond 560~GeV, at 95\% probability. Interestingly, the same constraints restrict the mass of a heavy charged Higgs to be below 226~GeV when the two neutral scalars are taken to be light.

While the measurements of the Higgs signal strengths constrain the mixing angle $\tilde\alpha$ to a very small value, the stability and pertubativity bounds force $\lambda_2$ and $\lambda_3$ to be in the range $[0,\, 9]$, and $\lambda_7$ to be in the range $[-2.5,\, 2.5]$. The mass of the charged Higgs plays an important role in constraining the parameter $\lambda_3$, whereas the mass of the new \cp-even scalar dictates the restriction on $\lambda_7$ through the effective cubic coupling $\lambda_{HH^+H^-}$ at one loop. Imposition of perturbativity at NLO level might restrict the scalar potential couplings in a more stringent fashion. Such an improvement of the code dedicated to the A2HDM in \HEPfit is currently being developed, and will be left for future analyses.

The alignment parameters are restricted by the constraints from Higgs signal strengths, flavour observables, and direct searches. 
Their correlations with the scalar mixing angle are governed by the Higgs signal strengths, while flavour observables control the correlations between the alignment parameters and the charged scalar mass. Since the dominant NP contributions to flavour observables
are proportional to quadratic powers of $\big(\varsigma_{u,d}/M_{H^\pm}\big)$, the scenarios with a heavier charged scalar permit larger values of $\varsigma_{u,d}$. On the other hand, the direct searches from the LHC provide restrictions to the combinations $\varsigma_u \varsigma_l$ and $\varsigma_d \varsigma_l$. Thus, larger values of $M_{H^\pm}$ indirectly restrict $\varsigma_l$ to lower values. Conversely, the scenarios with lighter $H^\pm$ require smaller values of $\varsigma_{u,d}$, in order to obey the flavour constraints, and that, in turn, weakens the direct search limits on $\varsigma_l$,
as \blue{illustrated by} the wider allowed ranges shown by this model parameter.

As discussed in Secs.~\ref{sec:slepton} and \ref{sec:susy}, the modes used by the LEP and LHC collaborations to look for the existence of sleptons exhibit a similar collider signature as those from probes of charged scalars decaying to $\tau^+\nu$, meaning that these searches can be taken into account while studying the phenomenology of the charged Higgs.
It is, once more, worth of note that one should only consider the data from left-handed sleptons, as they share the same weak isospin with the charged Higgs. Moreover, while sleptons exhibit a dominant decay to a particular lepton flavour, the branching fractions of other $H^\pm$ decay channels must be properly incorporated. With these considerations, we found that, among all the slepton searches, only the $\tilde \tau_L$ searches at the LHC have some influence on the  A2HDM global fits. This effect is, however, not very drastic: it imposes restrictions on the parameter space allowed to $\varsigma_l$ only for light charged scalar masses in the range of 115~GeV to 125~GeV.

Since the controversy about the SM prediction of $(g-2)_\mu$ has not been completely settled yet, we have not included this observable in our main global fits. Rather, we have presented, for the most favourable scenario with just a light charged scalar, the implications of two different theoretical estimates: that of the 2020 white paper, and the one derived from the latest BMW computations. As expected, the inclusion of the BMW results does not have any visible impact on the parameter space, as the lattice prediction of $a_\mu$ is now in perfect agreement with the experimental world average, whereas the result from the white paper tries to modify the $\varsigma_l$ vs $M_{H^\pm}$ correlation in such a way that the tension with the experimental value could be reduced with BSM scalar contributions. However, in terms of fit goodness, the inclusion of this $a_\mu^{\text{SM, WP}}$ from 2020 makes the resulting fit considerably worse.
Future consensus on the SM predictions, if they go in the direction of the BMW computations, would put to rest this necessity for the A2HDM to solve a \qt{$g-2$ anomaly}. Even an updated white paper value, depending on how such an endeavour would combine current estimations to produce the theoretical state-of-the-art, should already reduce the tension from the $5\sigma$ level it presently sits at.

To summarise, although the current constraints on light scalars are quite strong, our fits demonstrate that ample regions of the A2HDM parameter space are still allowed in the seven possible scenarios. This should motivate a renewed experimental effort in the search for light scalar or pseudoscalar bosons.

\section*{Acknowledgements}
We would like to acknowledge Milada M. M\"uhlleitner, Ulrich Nierste, Ayan Paul, Pablo Roig, Michael Spira, and Dominik St\"ockinger, for useful comments which helped improve some of the fits here presented.
A.P. would like to thank the high-energy physics group of the University of Granada for their hospitality during the time this article was being prepared.
This work has been supported
by MCIN/AEI/10.13039/501100011033 (grants PID2020-114473GB-I00, PID2023-146220NB-I00, and CEX2023-001292-S),
by Generalitat Valenciana (grant PROMETEO/2021/071),
by the European Research Council under the European Union’s Horizon 2020 research and innovation programme (grant agreement no. 949451),
and by a Royal Society University Research Fellowship (grant no. URF/R1/201553). 

\section*{Data Availability Statement}
This manuscript has no associated data or the data will not be deposited.

\section*{Code Availability Statement}
All the results in this paper can be reproduced using \HEPfit (at commit \href{https://github.com/silvest/HEPfit/tree/24ac2b10a2bf542b95a0d0d53a9d4d53f5ab0353}{24ac2b1}) --- following the procedures detailed in Ref.~\cite{DeBlas:2019ehy} --- together with configuration files containing the prior ranges given in Tab.~\ref{tab:prior}, examples of which may be provided upon request.

\clearpage

\FloatBarrier

\appendix

\section{Direct searches included}
\label{sec:direct_searches}
Here we provide the experimental direct searches that have been included in our analysis.
\subsection{Collider searches for light neutral scalars}
\label{sec:light_neutral}


\begin{table}[ht!]
	\centering{\small
		\begin{tabular}{|Q|l|c|c|c|}
			\hline\hline
			&Process & Range (GeV) & $\sqrt{s}$ (TeV) & Ref. \\
			\cline{2-5}\\[-5.7mm]\cline{2-5}
			\cellcolor{cms}&$pp \to A (b \, \bar b) \to \tau^+ \tau^- (b \, \bar b)$ &
			$25 < M_{A} < 80$ &\cellcolor{color8TeV} & 
			~\cite{CMS:2015qnd} \\
			\cline{2-3}\cline{5-5}
			&$pp \to h \to A A \to \tau^+ \tau^- \tau^+ \tau^-$ &
			$5 < M_{A} < 15$ & \cellcolor{color8TeV} &
			\multirow{3}{*}{~\cite{CMS:2017dmg}}  \\
			\cline{2-3}
			&$pp \to h \to A A \to \mu^+ \mu^- \, b \, \bar b$ &
			$25 < M_{A} < 62.5$ & \cellcolor{color8TeV} 8 &  \\
			\cline{2-3}
			&$pp \to h \to A A \to \mu^+ \mu^- \tau^+ \tau^-$ &
			$15 < M_{A} < 62.5$ & \cellcolor{color8TeV} &  \\
			\cline{2-3}\cline{5-5}
			&$pp \to A (b \, \bar b) \to \mu^+ \mu^- (b \, \bar b)$ &
			$25 < M_{A} < 60$ & \cellcolor{color8TeV} &
			~\cite{CMS:2017nmj} \\

            \cline{2-3}\cline{5-5}
			&$pp \to H \to \gamma \gamma$ &
			$80 < M_{H} < 110$ & \cellcolor{color8TeV} &
			~\cite{CMS:2018cyk} \\
			\cline{2-5}\\[-5.7mm]\cline{2-5}
			&$pp \to h \to A A \to \mu^+ \mu^- \tau^+ \tau^-$ &
			$15 < M_{A} < 62.5$ & \cellcolor{color13TeV} &
			~\cite{CMS:2018qvj} \\
			\cline{2-3}\cline{5-5}
			&$pp \to h \to A A \to \tau^+ \tau^- \, b \, \bar b$ &$12 < M_{A} < 60$ &  \cellcolor{color13TeV} & ~\cite{CMS:2024uru} \\
			\cline{2-3}\cline{5-5}
			&$pp \to h \to A A \to \mu^+ \mu^- \, b \, \bar b$ &$15 < M_{A} < 62$ &  \cellcolor{color13TeV} & ~\cite{CMS:2024uru} \\
			\cline{2-3}\cline{5-5}
			&$pp \to A (b \, \bar b) \to \tau^+ \tau^- (b \, \bar b)$ &
			$25 < M_{A} < 70$ &  \cellcolor{color13TeV} &
			~\cite{CMS:2019hvr} \\
			\cline{2-3}\cline{5-5}
			&$pp \to h \to A A \to \tau^+ \tau^- \tau^+ \tau^-$ &
			$4 < M_{A} < 15$ &  \cellcolor{color13TeV} &
			~\cite{CMS:2019spf} \\
			\cline{2-3}\cline{5-5}
			&$pp \to h \to \varphi_i^0 (Z) \to \ell^+ \ell^- (\ell^+ \ell^-)$ &
			\multirow{2}{*}{$4 < M_{\varphi_i^0} < 60$} &  \cellcolor{color13TeV}  &
			\multirow{2}{*}{~\cite{CMS:2021pcy}} \\
			&$pp \to h \to \varphi_i^0 \, \varphi_i^0  \to \ell^+ \ell^- \ell^+ \ell^-$ & &  \cellcolor{color13TeV} 13 &  \\
            
			\cline{2-3}\cline{5-5}
               &$pp \to h \to A Z \to \gamma \gamma \ell^+\ell^-$ &
			$1 < M_{A} < 30$ &  \cellcolor{color13TeV} &
			~\cite{CMS:2023eos} \\
			\cline{2-3}\cline{5-5}
            
			&$pp \to h \to A A \to \gamma \gamma \gamma \gamma$ &
			$15 < M_{A} < 62$ &  \cellcolor{color13TeV} &
			~\cite{CMS:2022xxa} \\
			\cline{2-3}\cline{5-5}
			&$pp \to h \to AA \to b\bar b b \bar b$ &
			$15 < M_{A} < 60$ &  \cellcolor{color13TeV} &
			~\cite{CMS:2024zfv} \\
			\cline{2-3}\cline{5-5}
			&$pp \to \varphi_i^0 (V)\to l^+l^- (l\nu/l^+l^-)$ &
			\multirow{2}{*}{$15 < M_{\varphi_i^0} < 350$} &  \cellcolor{color13TeV}&
			\multirow{2}{*}{~\cite{CMS:2024ulc}} \\
			\multirow{-17}{*}{\rotatebox{90}{\Large{CMS Searches}}}&$pp \to \varphi_i^0 (t\bar t) \to l^+l^- (t \bar t)$  & &  \cellcolor{color13TeV}&  \\

            \cline{2-3}\cline{5-5}
			&$pp \to H \to \gamma \gamma$ &
			$70 < M_{H} < 110$ & \cellcolor{color13TeV} &
			~\cite{CMS:2024yhz} \\ 
            \hline\hline
	\end{tabular}}
	\caption{Relevant CMS searches for a light neutral scalar.}
	\label{tab:CMS_light}
\end{table}


\begin{table}[ht!]
	\centering{\small
		\begin{tabular}{|R|l|c|c|c|}
			\hline\hline
			&Process & Range (GeV) & $\sqrt{s}$ (TeV) & Ref. \\
			\cline{2-5}\\[-5.7mm]\cline{2-5}
			&$pp \to h \to A A \to \tau^+ \tau^- \tau^+ \tau^-$ &
        $3.7 < M_{A} < 50$ &
        \cellcolor{color8TeV} &
        ~\cite{ATLAS:2015unc} \\
        \cline{2-3}\cline{5-5}
        &$pp \to h \to A A \to \gamma \gamma \gamma \gamma$ &
        $10 < M_{A} < 62$ & \cellcolor{color8TeV} \multirow{-2}{*}{8}&
        ~\cite{ATLAS:2015rsn} \\
        \cline{2-5}\\[-5.7mm]\cline{2-5}
        &$pp \to h \to \varphi_i^0\varphi_i^0 \to \gamma \gamma \, g g$ &
        $20 < M_{\varphi_i^0} < 60$ &
        \cellcolor{color13TeV} &
        ~\cite{ATLAS:2018jnf}  \\
        \cline{2-3}\cline{5-5}
        &$pp \to h (V) \to \varphi_i^0 \varphi_i^0 (\ell^+ \ell^-) \to b \, \bar b \, b \, \bar b \, (\ell^+ \ell^-)$ &
        $20 < M_{\varphi_i^0} < 60$ & \cellcolor{color13TeV} &
        ~\cite{ATLAS:2018pvw} \\
        \cline{2-3}\cline{5-5}
        &$pp \to h \to \varphi_i^0\varphi_i^0 \to \mu^+ \mu^- \, b \, \bar b$ &
        $20 < M_{\varphi_i^0} < 60$ & \cellcolor{color13TeV} &
        ~\cite{ATLAS:2018emt} \\
        \cline{2-3}\cline{5-5}
        &$pp \to h (Z) \to \varphi_i^0\varphi_i^0 (\ell^+ \ell^-) \to b \, \bar b \, b \, \bar b \, (\ell^+ \ell^-)$ &
        $15 < M_{\varphi_i^0} < 30$ & \cellcolor{color13TeV} &
        ~\cite{ATLAS:2020ahi} \\
        \cline{2-3}\cline{5-5}
        &$pp \to h \to A A \to \mu^+ \mu^- \, b \, \bar b$ &
        $16 < M_{A} < 62$ & \cellcolor{color13TeV} &
        ~\cite{ATLAS:2021hbr} \\
        \cline{2-3}\cline{5-5}
        &$pp \to h \to \varphi_i^0 (Z) \to \mu^+ \mu^- (\ell^+ \ell^-)$ &
        $15 < M_{\varphi_i^0} < 30$ & \cellcolor{color13TeV} &
        \multirow{2}{*}{~\cite{ATLAS:2021ldb}} \\
        &$pp \to h \to \varphi_i^0 \varphi_i^0 \to \mu^+ \mu^- \mu^+ \mu^-$ &
        $1 < M_{\varphi_i^0} < 60$ & \cellcolor{color13TeV} & \\        
        \cline{2-3}\cline{5-5}
        &$pp \to A (t \, \bar t) \to \mu^+ \mu^- (t \, \bar t)$ &
        $15 < M_{A} < 72$ & \cellcolor{color13TeV} &
        ~\cite{ATLAS:2023ofo} \\
        \cline{2-3}\cline{5-5}
        &$pp \to h \to A A \to b \, \bar b \,\tau^+\,\tau^-$ &
        $12 < M_{A} < 60$ & \cellcolor{color13TeV} &
        ~\cite{ATLAS:2024vpj} \\
        \cline{2-3}\cline{5-5}
        &$pp \to h \to A A \to \gamma\gamma\gamma\gamma$  & $10 < M_{A} < 62$& \cellcolor{color13TeV} &
        ~\cite{ATLAS:2023ian} \\
        \cline{2-3}\cline{5-5}
        \multirow{-14}{*}{\rotatebox{90}{\Large{ATLAS Searches}}}&$gg \to A \to \tau^+\,\tau^-$ &
        $20 < M_{A} < 90$ & \cellcolor{color13TeV} \multirow{-11}{*}{13}&
        ~\cite{ATLAS:2024rzd} \\

        \cline{2-3}\cline{5-5}
			&$pp \to H \to \gamma \gamma$ &
			$66 < M_{H} < 110$ & \cellcolor{color13TeV} &
			~\cite{ATLAS:2024bjr} \\ 
            
        \hline\hline
    \end{tabular}}
    \caption{Relevant ATLAS searches for a light neutral scalar.}
    \label{tab:ATLAS_light}
\end{table}


\begin{table}[ht!]
	\centering{\small
		\begin{tabular}{|S|l|c|c|c|}
			\hline\hline
			&Process & Range (GeV) & $\sqrt{s}$ (TeV) & Ref. \\
            \cline{2-5}\\[-5.7mm]\cline{2-5}
        &$e^+ e^- \to \varphi_i^{\rm even} \, Z \to \gamma \gamma \, Z$ &
        $20 < M_{\varphi_i^{\rm even}} < 116$ &
        0.088 -- 0.209 &
        {~\cite{LEPHiggsWorkingGroup:2001fac}}
        \\
        \cline{2-5}
        &$e^+ e^- \to \varphi_i^{\rm even} \, Z \to b \, \bar b \, Z$ &
        \multirow{2}{*}{$12 < M_{\varphi_i^{\rm even}} < 116$} & &
        \multirow{6}{*}{~\cite{ALEPH:2006tnd}}\\
        &$e^+ e^- \to \varphi_i^{\rm even} \, Z \to \tau^+ \tau^- \, Z$ & &
        \multirow{4}{*}{
        $\begin{matrix}
            0.091\\[-2mm]
            \text{to}\\[-2mm]
            0.209
        \end{matrix}$} &
        \\
        \cline{2-3}
        &$e^+ e^- \to \varphi_i^0 \, \varphi_j^0 \to b \, \bar b \, b \, \bar b$ &
        $\begin{matrix}
        15 < M_{\varphi_i^0} < 145 \\
        10 < M_{\varphi_j^0} < 105 \end{matrix}$ & &
        \\
        \cline{2-3}
        \multirow{-7}{*}{\rotatebox{90}{\Large{LEP Searches}}}&$e^+ e^- \to \varphi_i^0 \, \varphi_j^0 \to \tau^+ \tau^- \tau^+ \tau^-$ &
        $\begin{matrix}
        5 < M_{\varphi_i^0} < 150 \\
        5 < M_{\varphi_j^0} < 100 \end{matrix}$ & &
         \\
        \hline\hline
    \end{tabular}}
    \caption{Relevant LEP searches for a neutral scalar.}
    \label{tab:LEP_neutral}
\end{table}


\clearpage

\FloatBarrier

\subsection{Collider searches for heavy neutral scalars}
\label{sec:heavy_neutral}


\begin{table}[ht!]
	\centering{\small
		\begin{tabular}{|Q|l|c|c|c|}
			\hline\hline
			&Process & Range (TeV) & $\sqrt{s}$ (TeV) & Ref. \\
			\cline{2-5}\\[-5.7mm]\cline{2-5}
			\cellcolor{cms}&$pp\to  \varphi_i^0 \to hh \to (bb) (bb)$ &
			$  0.27 < M_{\varphi_i^0} < 1.1$ &\cellcolor{color8TeV}  & 
			~\cite{CMS:2015jal} \\
			\cline{2-3}\cline{5-5}

                &$pp\to  \varphi_i^0 \to hh \to (bb) (\gamma \gamma)$ &
			$0.26 < M_{\varphi_i^0} < 1.1$ & \cellcolor{color8TeV} &
			\cite{CMS:2016cma}  \\\cline{2-3}\cline{5-5}

			&$gg\to  \varphi_i^0 \to hh \to (bb) (\tau\tau)$ &
			$0.26 < M_{\varphi_i^0} < 0.35$ & \cellcolor{color8TeV} &
			\cite{CMS:2015uzk}  \\\cline{2-3}\cline{5-5}
   
                &$pp\to  \varphi_i^0 \to hh [\to (bb) (\tau\tau)]$ &
			$0.35 < M_{\varphi_i^0} < 1$ & \cellcolor{color8TeV}  &
			\cite{CMS:2017yfv}   \\\cline{2-3}\cline{5-5}

                &$gg\to  \varphi_i^0 \to hZ \to (bb) (\ell \ell)$ &
			$0.225 < M_{\varphi_i^0} < 0.6$ & \multirow{-2}{*}{ \cellcolor{color8TeV} 8} &
			\cite{CMS:2015flt}    \\\cline{2-3}\cline{5-5}

                &$gg\to  \varphi_i^0 \to hZ \to (\tau\tau) (\ell \ell)$ &
			$0.22 < M_{\varphi_i^0} < 0.35$ & \cellcolor{color8TeV} &
			\cite{CMS:2015uzk}    \\\cline{2-3}\cline{5-5}

                &$pp\to \varphi_3^0  \to \varphi_2^0 Z \to (bb) (\ell\ell)$ &
			$0.04 < M_{\varphi_i^0} < 1$ & \cellcolor{color8TeV} &
			\cite{CMS:2016xnc}     \\\cline{2-3}\cline{5-5}

                &$pp\to \varphi_3^0 \to \varphi_2^0 Z \to (\tau \tau) (\ell\ell)$ &
			$0.05 < M_{\varphi_i^0} < 1$ & \cellcolor{color8TeV} &
			\cite{CMS:2016xnc}     \\
			\cline{2-5}\\[-5.7mm]\cline{2-5}

			& \multirow{2}{*}{$pp \to  \varphi_i^0 \to hh \to (bb) (bb)$} &
			$0.26 < M_{\varphi_i^0} < 1.2$ & \cellcolor{color13TeV}  &
			~\cite{CMS:2018qmt} \\
			\cline{3-3}\cline{5-5}
   
                & &
			$1 < M_{\varphi_i^0} < 3$ & \cellcolor{color13TeV}  &
			~\cite{CMS:2024pjq} \\
			\cline{2-3}\cline{5-5}

                &$pp \to  \varphi_i^0 \to hh [\to (WW)(WW)/(WW)(\tau\tau)/(\tau\tau)(\tau\tau)]$ &
			$0.25 < M_{\varphi_i^0} < 1$ & \cellcolor{color13TeV}  &
			~\cite{CMS:2022kdx} \\
			\cline{2-3}\cline{5-5}

                &$pp \to  \varphi_i^0 \to hh \to (bb) (\gamma \gamma)$ &
			$0.26 < M_{\varphi_i^0} < 1$ & \cellcolor{color13TeV}  &
			~\cite{CMS:2023boe} \\
			\cline{2-3}\cline{5-5}

                &$pp \to  \varphi_i^0 \to hh \to (bb) (\tau \tau)$ &
			$0.25 < M_{\varphi_i^0} < 0.9 $ & \cellcolor{color13TeV}  &
			~\cite{CMS:2017hea} \\
			\cline{2-3}\cline{5-5}

                &$pp \to  \varphi_i^0 \to hh [\to (bb) (\tau \tau)]$ &
			$0.9 < M_{\varphi_i^0} < 4$ & \cellcolor{color13TeV}  &
			~\cite{CMS:2018kaz} \\
			\cline{2-3}\cline{5-5}

                &$pp \to  \varphi_i^0 \to hh \to (bb) (VV\to \ell \nu \ell \nu)$ &
			$0.26 < M_{\varphi_i^0} < 0.9$ & \cellcolor{color13TeV}  &
			~\cite{CMS:2017rpp} \\
			\cline{2-3}\cline{5-5}

                &$pp \to  \varphi_i^0 \to hh [\to (bb) (WW\to q \bar{q} \ell \nu)]$ &
			$0.8 < M_{\varphi_i^0} < 3.5$ & \cellcolor{color13TeV}  &
			~\cite{CMS:2019noi} \\
			\cline{2-3}\cline{5-5}

                &$pp \to  \varphi_i^0 \to hh \to (bb) [ZZ \to \ell \ell j j]$ &
			$0.26 < M_{\varphi_i^0} < 1$ & \cellcolor{color13TeV}  13&
			~\cite{CMS:2020jeo} \\
			\cline{2-3}\cline{5-5}
  
                &$pp \to  \varphi_i^0 \to hh \to (bb) [ZZ \to \ell \ell \nu \nu]$ &
			$0.26 < M_{\varphi_i^0} < 1$ & \cellcolor{color13TeV}  &
			~\cite{CMS:2020jeo} \\
			\cline{2-3}\cline{5-5}

                &\multirow{2}{*}{$pp \to  \varphi_i^0 \to hh [\to (bb) (WW/\tau\tau\to (q\bar{q}/\ell\nu) \ell\nu)]$} &
			$0.8 < M_{\varphi_i^0} < 4.5$ & \cellcolor{color13TeV}  &
			~\cite{CMS:2021roc} \\
			\cline{3-3}\cline{5-5}

                & &
			$0.25 < M_{\varphi_i^0} < 0.9$ & \cellcolor{color13TeV}  &
			~\cite{CMS:2024rgy} \\
			\cline{2-3}\cline{5-5}

                &\multirow{2}{*}{$gg\to  \varphi_i^0 \to (h\to b\bar{b})[Z\to \nu\bar{\nu}/\ell\bar{\ell}] $} &
			$0.22 < M_{\varphi_i^0} < 0.8$ & \cellcolor{color13TeV}  &
			~\cite{CMS:2019qcx} \\
			\cline{3-3}\cline{5-5}

                & &
			$0.8 < M_{\varphi_i^0} < 2$ & \cellcolor{color13TeV}  &
			~\cite{CMS:2018ljc} \\
			\cline{2-3}\cline{5-5}

                &$gg\to  \varphi_i^0 \to (h \to \tau\tau) (Z \to \ell \ell)$  &
			$0.22 < M_{\varphi_i^0} < 0.4$ & \cellcolor{color13TeV}  &
			~\cite{CMS:2019kca} \\
			\cline{2-3}\cline{5-5}

                &\multirow{2}{*}{$bb\to  \varphi_i^0 \to (h\to b\bar{b})[Z\to \nu\bar{\nu}/\ell\bar{\ell}] $}  &
			$0.22 < M_{\varphi_i^0} < 0.8$ & \cellcolor{color13TeV}  &
			~\cite{CMS:2019qcx} \\
			\cline{3-3}\cline{5-5}

                \multirow{-26}{*}{\rotatebox{90}{\Large{CMS Searches}}} &  &
			$0.8 < M_{\varphi_i^0} < 2$ & \cellcolor{color13TeV}  &
			~\cite{CMS:2018ljc} \\
			
			\hline\hline
	\end{tabular}}
	\caption{Direct searches in CMS for neutral heavy scalars, $\varphi_i^0 = H, A$, with final states including the Higgs boson or other neutral scalars. $\varphi_3$ denotes the heaviest scalar,  $V = W,Z$, $\ell = e, \mu$. The parenthesis show the final decay of the SM particles produced from the NP particles. The square brackets are used when the values of \sigbr are shown in terms of the primary decay (i.e. the NP particle decay) but a particular decay channel of the SM particles is used to obtain those values.}
	\label{tab:CMS_heavy_scalar_to_scalar}
\end{table}


\begin{table}[ht!]
	\centering{\small
		\begin{tabular}{|R|l|c|c|c|}
			\hline\hline
			\cellcolor{atlas}&Process & Range (TeV) & $\sqrt{s}$ (TeV) & Ref. \\
			\cline{2-5}\\[-5.7mm]\cline{2-5}
   
			\cellcolor{atlas}&$gg\to \varphi_i^0 \to hh$ &
			$  0.26 < M_{\varphi_i^0} < 1$ &\cellcolor{color8TeV}  & 
			~\cite{ATLAS:2015sxd} \\
			\cline{2-3}\cline{5-5}

                \cellcolor{atlas}&$gg\to  \varphi_i^0 \to hZ \to (bb) Z$ &
			$0.22 < M_{\varphi_i^0} < 1$ & \cellcolor{color8TeV} 8 &
			\cite{ATLAS:2015kpj}  \\\cline{2-3}\cline{5-5}

                \cellcolor{atlas}&$gg\to  \varphi_i^0 \to hZ \to (\tau\tau) Z$ &
			$0.05 < M_{\varphi_i^0} < 1$ & \cellcolor{color8TeV} &
			\cite{ATLAS:2015kpj}     \\
			\cline{2-5}\\[-5.7mm]\cline{2-5}

                &$pp \to  \varphi_i^0 \to hh [\to (bb) (bb)]$ &
			$0.251 < M_{\varphi_i^0} < 5$ & \cellcolor{color13TeV}  &
			~\cite{ATLAS:2022hwc} \\
			\cline{2-3}\cline{5-5}

                &$pp \to  \varphi_i^0 \to hh [\to (bb) (\gamma \gamma)]$ &
			$0.251 < M_{\varphi_i^0} < 1$ & \cellcolor{color13TeV}  &
			~\cite{ATLAS:2021ifb}  \\
			\cline{2-3}\cline{5-5}

                &\multirow{2}{*}{$pp \to \varphi_i^0 \to hh [\to (bb) (\tau \tau)]$} &
			$0.251 < M_{\varphi_i^0} < 1.6$ & \cellcolor{color13TeV}  &
			~\cite{ATLAS:2022xzm}  \\
			\cline{3-3}\cline{5-5}

                & &
			$1 < M_{\varphi_i^0} < 3$ & \cellcolor{color13TeV}  &
			~\cite{ATLAS:2020azv}  \\
			\cline{2-3}\cline{5-5}

                &$pp \to  \varphi_i^0 \to hh [\to (bb) (WW)]$ &
			$0.5 < M_{\varphi_i^0} < 3$ & \cellcolor{color13TeV}  &
			~\cite{ATLAS:2018fpd}  \\
			\cline{2-3}\cline{5-5}

                &$gg \to  \varphi_i^0 \to hh \to (\gamma \gamma) (WW)$ &
			$0.26 < M_{\varphi_i^0} < 0.5$ & \cellcolor{color13TeV} 13 &
			~\cite{ATLAS:2018hqk}  \\
			\cline{2-3}\cline{5-5}

                &$gg\to  \varphi_i^0 \to hZ [\to (bb) Z]$&
			\multirow{2}{*}{$0.22 < M_{\varphi_i^0} < 2$} & \cellcolor{color13TeV}  &
			\multirow{2}{*}{~\cite{ATLAS:2022enb}}  \\
			\cline{2-2}

                &$bb\to  \varphi_i^0 \to hZ [\to (bb) Z]$&
			 & \cellcolor{color13TeV}  &
			  \\
			\cline{2-3}\cline{5-5}

                &$gg\to \varphi_{3}^0 \to \varphi_2^0 Z \to (bb) Z$&
			$0.13 < M_{\varphi_3^0} < 0.7$ & \cellcolor{color13TeV}  &
			\multirow{3}{*}{~\cite{ATLAS:2020gxx} } \\
			\cline{2-2}

                &$bb\to \varphi_{3}^0 \to \varphi_2^0 Z \to (bb) Z$&
			  $0.23 < M_{\varphi_2^0} < 0.8$ & \cellcolor{color13TeV}  &
			  \\
			\cline{2-3}

                \multirow{-15}{*}{\rotatebox{90}{ \cellcolor{atlas} \Large{ATLAS Searches}}} &$gg\to \varphi_{3}^0 \to \varphi_2^0 Z \to (WW) Z$   &
			$0.2 < M_{\varphi_3^0} < 0.7$ / $0.3 < M_{\varphi_2^0} < 0.8$ & \cellcolor{color13TeV}  &
			 \\
			\hline\hline
	\end{tabular}}
	\caption{Direct searches in ATLAS for neutral heavy scalars, $\varphi_i^0 = H, A$, with final states including the Higgs boson or other neutral scalars. $\varphi_3$ denotes the heaviest scalar,  $V = W,Z$, $\ell = e, \mu$. The parenthesis show the final decay of the SM particles produced from the NP particles. The square brackets are used when the values of \sigbr are shown in terms of the primary decay (i.e. the NP particle decay) but a particular decay channel of the SM particles is used to obtain those values.}
	\label{tab:ATLAS_heavy_scalar_to_scalar}
\end{table}


\begin{table}[ht!]
	\centering{\small
		\begin{tabular}{|Q|l|c|c|c|}
			\hline\hline
			&Process & Range (TeV) & $\sqrt{s}$ (TeV) & Ref. \\
			\cline{2-5}\\[-5.7mm]\cline{2-5}

                &$pp\to \varphi_i^0 \to Z\gamma \to (\ell \ell) \gamma$&
			$0.2 < M_{\varphi_i^0} < 1.2$ & \cellcolor{color8TeV} &
			\cite{CMS:2016all}    \\\cline{2-3}\cline{5-5}
   
                &$pp \to \varphi_i^0\to VV$&
			$0.145 < M_{\varphi_i^0} < 1$ & \multirow{-2}{*}{ \cellcolor{color8TeV} 8}&
			\cite{CMS:2015hra}    \\
			\cline{2-5}\\[-5.7mm]\cline{2-5}

                &$gg \to \varphi_i^0\to \gamma \gamma$ &
			$0.6 < M_{\varphi_i^0} < 5$ & \cellcolor{color13TeV}  &
			~\cite{CMS:2024nht} \\
			\cline{2-3}\cline{5-5}

                 &$pp \to \varphi_i^0\to Z \gamma [\to (\ell \ell \,\&\, qq) \gamma ]$  &
			$0.35 < M_{\varphi_i^0} < 4$ & \cellcolor{color13TeV}  &
			~\cite{CMS:2017dyb}\\
                \cline{2-3}\cline{5-5}

			&$pp\to \varphi_i^0\to ZZ [\to (\ell \ell) (qq,\nu\nu,\ell\ell)]$ &
			$0.13 < M_{\varphi_i^0} < 3$ & \cellcolor{color13TeV}  &
			~\cite{CMS:2018amk} \\
			\cline{2-3}\cline{5-5}

                & $pp\to \varphi_i^0\to ZZ [\to (qq)(\nu\nu)]$ &
			$1 < M_{\varphi_i^0} < 4$ & \cellcolor{color13TeV}  &
			~\cite{CMS:2018ygj}\\
			\cline{2-3}\cline{5-5}

                & $VV \to \varphi_i^0\to WW$ &
			\multirow{2}{*}{$0.2 < M_{\varphi_i^0} < 3$} & \cellcolor{color13TeV}     &\multirow{2}{*}{~\cite{CMS:2019bnu}}\\
			\cline{2-2}

                & $gg \to \varphi_i^0\to WW$ &
			  & \multirow{-2}{*}{\cellcolor{color13TeV} 13}  &
			\\\cline{2-3}\cline{5-5}

                & $pp\to \varphi_i^0\to WW[\to (\ell \nu) (qq)]$ &
			$1 < M_{\varphi_i^0} < 4.4$ & \multirow{1}{*}{\cellcolor{color13TeV} }  &
			~\cite{CMS:2018dff}\\
			\cline{2-3}\cline{5-5}

                & $gg \to \varphi_i^0\to WW$ &
			\multirow{2}{*}{$1 < M_{\varphi_i^0} < 4.5$} & \cellcolor{color13TeV}     &\multirow{2}{*}{~\cite{CMS:2021klu}}\\
			\cline{2-2}

                & $VV \to \varphi_i^0\to WW$ &
			  & \cellcolor{color13TeV}  &
			\\\cline{2-3}\cline{5-5}
   
                \multirow{-10}{*}{\rotatebox{90}{\Large{CMS Searches}}} &$(gg\!+\!VV)\to \varphi_i^0\to WW \to (\ell \nu) (\ell \nu)$ &
			$0.2 < M_{\varphi_i^0} < 1$ & \cellcolor{color13TeV}  &
			~\cite{CMS:2016jpd}\\
			
			\hline\hline
	\end{tabular}}
	\caption{Direct searches in CMS for neutral heavy scalars, $\varphi_i^0 = H, A$, with vector-boson final states. $V = W,Z$, $\ell = e, \mu$. The parenthesis show the final decay of the SM particles produced from the NP particles. The square brackets are used when the values of \sigbr are shown in terms of the primary decay (i.e. the NP particle decay) but a particular decay channel of the SM particles is used to obtain those values.}
	\label{tab:CMS_heavy_scalar_to_bosons}
\end{table}


\begin{table}[ht!]
	\centering{\small
		\begin{tabular}{|R|l|c|c|c|}
			\hline\hline
			\cellcolor{atlas}&Process & Range (TeV) & $\sqrt{s}$ (TeV) & Ref. \\
			\cline{2-5}\\[-5.7mm]\cline{2-5}

                \cellcolor{atlas}&$gg\to \varphi_i^0 \to \gamma\gamma$ &
			$  0.065 < M_{\varphi_i^0} < 0.6$ &\multirow{-2}{*}{\cellcolor{color8TeV}  }& 
			~\cite{ATLAS:2014jdv} \\
			\cline{2-3}\cline{5-5}

                \cellcolor{atlas}&$pp\to \varphi_i^0 \to Z\gamma \to (\ell \ell) \gamma$ &
			$0.2 < M_{\varphi_i^0} < 1.6$ & \cellcolor{color8TeV} &
			\cite{ATLAS:2014lfk}     \\   
			\cline{2-3}\cline{5-5}
                
                \cellcolor{atlas}&$gg\to \varphi_i^0\to ZZ$ &
			$  0.14 < M_{\varphi_i^0} < 1$ &\multirow{1}{*}{\cellcolor{color8TeV} 8}  & 
			~\cite{ATLAS:2015pre} \\
			\cline{2-3}\cline{5-5}

                \cellcolor{atlas}&$gg\to \varphi_i^0\to WW$&
			\multirow{2}{*}{$  0.3 < M_{\varphi_i^0} < 1.5$} 
                &\multirow{1}{*}{\cellcolor{color8TeV} }  & 
			\multirow{2}{*}{~\cite{ATLAS:2015iie}} \\
			\cline{2-2}

                \cellcolor{atlas}&$VV \to \varphi_i^0\to WW$ &
			  & \cellcolor{color8TeV}  &
			  \\ \cline{2-5}\\[-5.7mm]\cline{2-5}

                \cellcolor{atlas}&$pp \to \varphi_i^0\to \gamma \gamma$ &
			$  0.15 < M_{\varphi_i^0} < 3$ &\cellcolor{color13TeV}  & 
			~\cite{ATLAS:2021uiz} \\
			\cline{2-3}\cline{5-5}

                \cellcolor{atlas}&$gg \to \varphi_i^0\to Z \gamma [\to (\ell \ell) \gamma ]$ &
			$  0.22 < M_{\varphi_i^0} < 3.4$ &\cellcolor{color13TeV}  & 
			~\cite{ATLAS:2023wqy} \\
			\cline{2-3}\cline{5-5}

                \cellcolor{atlas}&$gg \to \varphi_i^0\to Z \gamma [\to (qq) \gamma ]$  &$  1.02 < M_{\varphi_i^0} < 6.8$ & \cellcolor{color13TeV}  &
			~\cite{ATLAS:2023kcu} \\
			\cline{2-3}\cline{5-5}

                \cellcolor{atlas}& $gg\to \varphi_i^0 \to ZZ [\to (\ell \ell) (\ell \ell, \nu \nu)]$ &
			\multirow{2}{*}{$  0.2 < M_{\varphi_i^0} < 2$} &\cellcolor{color13TeV}  & 
			\multirow{2}{*}{~\cite{ATLAS:2020tlo}} \\
			\cline{2-2}

                \cellcolor{atlas}& $VV\to \varphi_i^0\to ZZ [\to (\ell \ell) (\ell \ell, \nu \nu)]$ &
			 &\cellcolor{color13TeV}  & 
			 \\	\cline{2-3}\cline{5-5}

                \cellcolor{atlas}& $gg\to \varphi_i^0\to ZZ [\to (\ell \ell, \nu \nu) (qq)]$ & \multirow{2}{*}{$  0.3 < M_{\varphi_i^0} < 3$} &\cellcolor{color13TeV}  & 
			\multirow{2}{*}{~\cite{ATLAS:2017otj}} \\
			\cline{2-2}

                \cellcolor{atlas}& $VV\to \varphi_i^0\to ZZ [\to (\ell \ell, \nu \nu) (qq)]$  &  &\cellcolor{color13TeV}   & 
			  \\
			\cline{2-3}\cline{5-5}

                \cellcolor{atlas}& $gg\to \varphi_i^0\to WW [\to (e \nu) (\mu \nu)]$   & \multirow{2}{*}{$ 0.2 < M_{\varphi_i^0} < 4$} &\multirow{-2}{*}{\cellcolor{color13TeV} 13}  & 
			\multirow{2}{*}{~\cite{ATLAS:2022qlc}} \\
			\cline{2-2}

                \cellcolor{atlas}& $VV\to \varphi_i^0\to WW [\to (e \nu) (\mu \nu)]$   &  &\cellcolor{color13TeV} & 
			\\	\cline{2-3}\cline{5-5}

                \cellcolor{atlas}& $gg\to \varphi_i^0\to WW[\to (\ell \nu) (qq)]$   & \multirow{2}{*}{$0.3 < M_{\varphi_i^0} < 3$} &\cellcolor{color13TeV}  & \multirow{2}{*}{~\cite{ATLAS:2017jag}} \\
			\cline{2-2}

                \cellcolor{atlas}&  $VV\to \varphi_i^0\to WW[\to (\ell \nu) (qq)]$   &  &\cellcolor{color13TeV}  & 
			  \\\cline{2-3}\cline{5-5}

                \cellcolor{atlas}&  $pp\to \varphi_i^0\to VV [\to (qq) (qq)]$  & $  1.2 < M_{\varphi_i^0} < 3$ &\cellcolor{color13TeV}  & 
			~\cite{ATLAS:2017zuf} \\
			\cline{2-3}\cline{5-5}

                \cellcolor{atlas}&  $gg \to \varphi_i^0\to VV$  & \multirow{2}{*}{$  0.2 < M_{\varphi_i^0} < 5.2$} &\cellcolor{color13TeV}  & 
			\multirow{2}{*}{~\cite{ATLAS:2020fry}}\\
			\cline{2-2}

                \multirow{-20}{*}{\rotatebox{90}{ \cellcolor{atlas} \Large{ATLAS Searches}}} &  $VV \to \varphi_i^0\to VV$  &
			  & \cellcolor{color13TeV}  &
			  \\
			
			\hline\hline
	\end{tabular}}
	\caption{Direct searches in ATLAS for neutral heavy scalars, $\varphi_i^0 = H, A$, with vector-boson final states. $V = W,Z$, $\ell = e, \mu$. The parenthesis show the final decay of the SM particles produced from the NP particles. The square brackets are used when the values of \sigbr are shown in terms of the primary decay (i.e. the NP particle decay) but a particular decay channel of the SM particles is used to obtain those values.}
	\label{tab:ATLAS_heavy_scalar_to_bosons}
\end{table}


\begin{table}[ht!]
	\centering{\small
		\begin{tabular}{|Q|l|c|c|c|}
			\hline\hline
			&Process & Range (TeV) & $\sqrt{s}$ (TeV) & Ref. \\
			\cline{2-5}\\[-5.7mm]\cline{2-5}

                \cellcolor{cms}&$bb \to \varphi_i^0 \to bb$ &
			$  0.1 < M_{\varphi_i^0} < 0.9$ &\cellcolor{color8TeV}  &			~\cite{CMS:2015grx} \\
			\cline{2-3}\cline{5-5}

                &$gg \to \varphi_i^0\to bb$ &
			$0.33 < M_{\varphi_i^0} < 1.2$ & \cellcolor{color8TeV} &
			~\cite{CMS:2018kcg}    \\\cline{2-3}\cline{5-5}

                &$bb\to \varphi_i^0 \to \mu\mu$ &
			\multirow{2}{*}{$0.12 < M_{\varphi_i^0} < 0.5$} & \cellcolor{color8TeV} &
			\multirow{2}{*}{~\cite{CMS:2015ooa} }   \\\cline{2-2}

                &$gg\to \varphi_i^0 \to \mu\mu$&
			  & \multirow{-2}{*}{\cellcolor{color8TeV} 8} &
			   \\\cline{2-3}\cline{5-5}

                &$bb\to \varphi_i^0 \to \tau\tau$ &
			\multirow{2}{*}{$0.09 < M_{\varphi_i^0} < 1$} & \cellcolor{color8TeV}     & \multirow{2}{*}{~\cite{CMS:2015mca} }   \\\cline{2-2}

                &$gg\to \varphi_i^0 \to \tau\tau$&
			  & \cellcolor{color8TeV} &
			  \\\cline{2-5}\\[-5.7mm]\cline{2-5}

			&$tt/tW/tq \to \varphi_i^0 \to tt$ &
			$0.35 < M_{\varphi_i^0} < 0.65$ & \cellcolor{color13TeV}  &
			~\cite{CMS:2019rvj} \\
			\cline{2-3}\cline{5-5}

                &\multirow{2}{*}{$pp \to \varphi_i^0\to bb$} &
			$0.55 < M_{\varphi_i^0} < 1.2$ & \cellcolor{color13TeV}  &
			~\cite{CMS:2016ncz} \\
			\cline{3-3}\cline{5-5}

                & &$0.05 < M_{\varphi_i^0} < 0.35$ & \cellcolor{color13TeV}  &
			~\cite{CMS:2018pwl} \\
			\cline{2-3}\cline{5-5}

                &$bb \to \varphi_i^0\to bb$ &
			$0.3 < M_{\varphi_i^0} < 1.3$ & \cellcolor{color13TeV}  &
			~\cite{CMS:2018hir} \\
			\cline{2-3}\cline{5-5}

                &$bb\to \varphi_i^0 \to \mu\mu$ &
			\multirow{2}{*}{$0.14 < M_{\varphi_i^0} < 1$} & \multirow{-2}{*}{\cellcolor{color13TeV} 13}  &
			\multirow{2}{*}{~\cite{CMS:2019mij}} \\
			\cline{2-2}

                &$gg\to \varphi_i^0 \to \mu\mu$ &
			  & \multirow{1}{*}{\cellcolor{color13TeV} }  &
			  \\\cline{2-3}\cline{5-5}

                &$bb\to \varphi_i^0 \to \tau \tau$ &
			\multirow{2}{*}{$0.06 < M_{\varphi_i^0} < 3.5$} & \cellcolor{color13TeV}   &
			\multirow{2}{*}{~\cite{CMS:2022goy}} \\
			\cline{2-2}

                \multirow{-15}{*}{\rotatebox{90}{\Large{CMS Searches}}} &$gg\to \varphi_i^0 \to \tau \tau$ &
			  & \cellcolor{color13TeV}  &
			  \\ 			
			\hline\hline
	\end{tabular}}
	\caption{Direct searches in CMS for neutral heavy scalars, $\varphi_i^0 = H, A$, with quarks and leptons ($\ell= e,\mu$) final states.}
	\label{tab:CMS_heavy_scalar_to_fermions}
\end{table}


\begin{table}[ht!]
	\centering{\small
		\begin{tabular}{|R|l|c|c|c|}
			\hline\hline
			\cellcolor{atlas}&Process & Range (TeV) & $\sqrt{s}$ (TeV) & Ref. \\
			\cline{2-5}\\[-5.7mm]\cline{2-5}

			\cellcolor{atlas}&$gg\to \varphi_i^0 \to \tau\tau$ &
			 \multirow{2}{*}{$  0.09 < M_{\varphi_i^0} < 1$} &\cellcolor{color8TeV}  & 
			\multirow{2}{*}{~\cite{ATLAS:2014vhc}} \\
			\cline{2-2}

                \cellcolor{atlas}&$bb\to \varphi_i^0 \to \tau\tau$ &
			  &\multirow{-2}{*}{\cellcolor{color8TeV} 8} & 
			\\   
			\cline{2-5}\\[-5.7mm]\cline{2-5}

                \cellcolor{atlas}&$bb \to \varphi_i^0 \to tt$ &\multirow{2}{*}{
			$  0.4 < M_{\varphi_i^0} < 1$} &\cellcolor{color13TeV}  & 
			~\cite{ATLAS:2018cye} \\
			\cline{2-2}\cline{5-5}

                \cellcolor{atlas}& $tt \to \varphi_i^0 \to tt$ &
			  &\cellcolor{color13TeV}  & 
			~\cite{ATLAS:2022rws} \\
			\cline{2-3}\cline{5-5}

                \cellcolor{atlas}&$pp \to \varphi_i^0\to bb$  ($\geq 1$ b-jet) &
			$  1.4 < M_{\varphi_i^0} < 6.6$ &\cellcolor{color13TeV}  & 
			\multirow{3}{*}{~\cite{ATLAS:2018tfk}} \\
			\cline{2-3}
   
                \cellcolor{atlas}&\multirow{2}{*}{$pp \to \varphi_i^0\to bb$} &
			$  0.6 < M_{\varphi_i^0} < 1.25$ &\cellcolor{color13TeV}  & 
			  \\\cline{3-3}
				
                \cellcolor{atlas}&  &
			$  1.25 < M_{\varphi_i^0} < 6.2$ &\cellcolor{color13TeV}  & 
			\\\cline{2-3}\cline{5-5}
    
                \cellcolor{atlas}&$bb \to \varphi_i^0\to bb$ &
			$  0.45 < M_{\varphi_i^0} < 1.4$ &\multirow{-2}{*}{\cellcolor{color13TeV} 13} & 
			~\cite{ATLAS:2019tpq} \\
			\cline{2-3}\cline{5-5}

                \cellcolor{atlas}&$bb\to \varphi_i^0 \to \mu\mu$ &
			\multirow{2}{*}{$  0.2 < M_{\varphi_i^0} < 1$}       
                &\cellcolor{color13TeV}  & 
			\multirow{2}{*}{~\cite{ATLAS:2019odt}} \\
			\cline{2-2}

                \cellcolor{atlas}&$gg\to \varphi_i^0 \to \mu\mu$ &
			  &\cellcolor{color13TeV}  & 
			  \\\cline{2-3}\cline{5-5}

                \cellcolor{atlas}&$bb\to \varphi_i^0 \to \tau\tau$ &
			\multirow{2}{*}{$  0.2 < M_{\varphi_i^0} < 2.5$} &\cellcolor{color13TeV}  & 
			\multirow{2}{*}{~\cite{ATLAS:2020zms}} \\
			\cline{2-2}

                \multirow{-12}{*}{\rotatebox{90}{ \cellcolor{atlas} \Large{ATLAS Searches}}}&$gg\to \varphi_i^0 \to \tau\tau$ &
			  &\cellcolor{color13TeV}  & 
			  \\\hline\hline
	\end{tabular}}
	\caption{Direct searches in ATLAS for neutral heavy scalars, $\varphi_i^0 = H, A$, with quarks and leptons ($\ell= e,\mu$) final states.}
	\label{tab:ATLAS_heavy_scalar_to_fermions}
\end{table}


\FloatBarrier

\subsection{Collider searches for charged scalars}
\label{sec:charged}


\begin{table}[ht!]
	\centering{\small
		\begin{tabular}{|Q|l|c|c|c|}
			\hline\hline
			&Process & Range (GeV) & $\sqrt{s}$ (TeV) & Ref.  \\
	\cline{2-5}\\[-5.7mm]\cline{2-5}
        &$t \to (H^+ \to \tau^+ \nu) \, b$ &
        $80 < M_{H^+} < 160$ & \cellcolor{color8TeV} &
        ~\cite{CMS:2015lsf}  \\
        \cline{2-3}\cline{5-5}
        &$t \to (H^+ \to c \, \bar{s} = 100\%) \, b$ &
        $90 < M_{H^+} < 160$ & \cellcolor{color8TeV} &
        ~\cite{CMS:2015yvc}  \\
        \cline{2-3}\cline{5-5}
        &$t \to (H^+ \to c \, \bar{b} = 100\%) \, b$ &
        $90 < M_{H^+} < 150$ & \cellcolor{color8TeV} \multirow{-3}{*}{8}&
        ~\cite{CMS:2018dzl} \\
        \cline{2-5}\\[-5.7mm]\cline{2-5}
        &$t \to b \, (H^+ \to W^+ A) \to b \, (W^+ \mu^+ \mu^-)$ & $\begin{matrix}
        15 < M_A < 75 \\
        100 < M_{H^+} < 160 \end{matrix}$ &
        \multirow{2}{*}{13} \cellcolor{color13TeV}&
        ~\cite{CMS:2019idx} \\
        \cline{2-3}\cline{5-5}
        \multirow{-7}{*}{\rotatebox{90}{\Large{CMS Searches}}}&$t \to (H^+ \to c \, \bar{s} = 100\%) \, b$ &
        $80 < M_{H^+} < 160$ & \cellcolor{color13TeV}&
        ~\cite{CMS:2020osd} \\
        \hline\hline
    \end{tabular}}
    \caption{Relevant CMS searches for a light charged scalar.}
    \label{tab:CMS_charged}
\end{table}


\begin{table}[ht!]
	\centering{\small
		\begin{tabular}{|R|l|c|c|c|}
			\hline\hline
			&Process & Range (GeV) & $\sqrt{s}$ (TeV) & Ref.  \\
	\cline{2-5}\\[-5.7mm]\cline{2-5}
        &$t \to (H^+ \to \tau^+ \nu) \, b$ &
        $80 < M_{H^+} < 160$ &
        \cellcolor{color8TeV} 8 &
        ~\cite{ATLAS:2014otc} \\
        \cline{2-5}\\[-5.7mm]\cline{2-5}
        &$t \to (H^+ \to c \, \bar{b}) \, b$ &
        $60 < M_{H^+} < 160$ &
        \cellcolor{color13TeV}&
       ~\cite{ATLAS:2023bzb} \\
        \cline{2-3}\cline{5-5}
        &$t \to b \, (H^+ \to W^+ A) \to b \, (W^+ \mu^+ \mu^-)$ & $\begin{matrix}
        15 < M_A < 72 \\
        120 < M_{H^+} < 160 \end{matrix}$ & \cellcolor{color13TeV}&
        ~\cite{ATLAS:2023ofo} \\
        \cline{2-3}\cline{5-5}
        \multirow{-6}{*}{\rotatebox{90}{\Large{ATLAS Searches}}}&$t \to (H^+ \to c\,\bar{s})\,b$  & $60< M_{H^+}< 168$ &  \cellcolor{color13TeV}\multirow{-4}{*}{13} & \cite{ATLAS:2024oqu} \\
        \hline\hline
    \end{tabular}}
    \caption{Relevant ATLAS searches for a light charged scalar.}
    \label{tab:ATLAS_charged}
\end{table}


\begin{table}[ht!]
	\centering{\small
		\begin{tabular}{|S|l|c|c|c|}
			\hline\hline
			&Process & Range (GeV) & $\sqrt{s}$ (TeV) & Ref.  \\
	\cline{2-5}\\[-5.7mm]\cline{2-5}
        &$e^+ e^- \to H^+ H^- \to \tau^+ \nu \, \tau^- \bar\nu$ &
        \multirow{2}{*}{$43 < M_{H^+} < 95$} & \multirow{5}{*}{
        $\begin{matrix}
            0.183\\[-2mm]
            \text{to}\\[-2mm]
            0.209
        \end{matrix}$}
         &
        \multirow{2}{*}{~\cite{ALEPH:2013htx}} \\
        &$e^+ e^- \to H^+ H^- \to q \, \bar{q} \, q \, \bar{q}$ & & &
         \\
        \cline{2-3}\cline{5-5}
        &$e^+ e^- \to H^+ H^- \to q \, \bar{q} \, \tau^- \bar\nu$ &
        $50 < M_{H^+} < 93$ &
         &
        \multirow{3}{*}{~\cite{OPAL:2008tdn}} \\
        \cline{2-3}
        &$e^+ e^- \to H^+ H^- \to (A \to b \, \bar{b}) W^{+*} (A \to b \, \bar{b}) W^{-*}$ &
        $12 < M_A < M_{H^+}$ & &
         \\
        \multirow{-6}{*}{\rotatebox{90}{\Large{LEP Searches}}}&$e^+ e^- \to H^+ H^- \to (A \to b \, \bar{b}) W^{+*} \tau^- \bar\nu$ &
        $40 < M_{H^+} < 94$ & &
         \\
        \hline\hline
    \end{tabular}}
    \caption{Relevant LEP searches for a light charged scalar.}
    \label{tab:LEP_charged}
\end{table}


\clearpage

\begin{table}[ht!]
	\centering{\small
 		\begin{tabular}{|c|l|c|c|c|}
                \cline{2-5}
                \multicolumn{1}{c}{\vspace{-5.7mm}}\\
                \cline{2-5}
			\multicolumn{1}{c|}{}&Process & Range (TeV) & $\sqrt{s}$ (TeV) & Ref. \\\hline\hline

        \cellcolor{cms} \begin{tikzpicture}[overlay]
        \fill[cms] (-0.44, 0.2) rectangle (0.33, -1.0);
        \end{tikzpicture}& $pp\to H^{ \pm} \to \tau^{ \pm} \nu $ &
			\multirow{2}{*}{$0.18 < M_{H^\pm} < 0.6$} &\cellcolor{color8TeV}  & 
			\multirow{2}{*}{~\cite{CMS:2015lsf}} \\
			\cline{2-2}

                \cellcolor{cms}&$pp\to H^{ \pm} \to t b $ &
			  & \multirow{-2}{*}{\cellcolor{color8TeV} 8} &
			    \\   
			\cline{2-5}\\[-5.7mm]\cline{2-5}

                \cellcolor{cms}& $pp\to H^{ \pm} \to \tau^{ \pm} \nu $ &
			$  0.08 < M_{H^{\pm}} < 3$ &\cellcolor{color13TeV}  & 
			~\cite{CMS:2019bfg} \\
			\cline{2-3}\cline{5-5}

                \multirow{-4}{*}{\rotatebox{90}{ \cellcolor{cms} \Large{CMS }}} &  $pp\to H^{ \pm} \to t b $  &
			$  0.2 < M_{H^{\pm}} < 3 $ & \multirow{-2}{*}{\cellcolor{color13TeV} 13}  &
			~\cite{CMS:2020imj} \\\hline\hline

                \cellcolor{atlas} \begin{tikzpicture}[overlay]
        \fill[atlas] (-0.44, 0.2) rectangle (0.33, -1.0);
    \end{tikzpicture}& $pp\to H^\pm \to \tau^\pm \nu $ &
			$  0.18 < M_{H^\pm} < 1$ &\cellcolor{color8TeV}  & 
			~\cite{ATLAS:2014otc} \\
			\cline{2-3}\cline{5-5}

                \cellcolor{atlas}& $pp\to H^\pm \to t b $ &
			$0.2 < M_{H^\pm} < 0.6$ & \multirow{-2}{*}{\cellcolor{color8TeV} 8}  &
			\cite{ATLAS:2015nkq}     \\   
			\cline{2-5}\\[-5.7mm]\cline{2-5}

                \cellcolor{atlas}& $pp\to H^{\pm} \to \tau^\pm \nu $ &
			$  0.09 < M_{H^\pm} < 2$ & \cellcolor{color13TeV}  & 
			~\cite{ATLAS:2018gfm} \\
			\cline{2-3}\cline{5-5}

                \multirow{-4}{*}{\rotatebox{90}{ \cellcolor{atlas} \Large{ATLAS }}} &  $pp\to H^\pm \to tb $  &
			$  0.2 < M_{H^\pm}  < 2$ & \multirow{-2}{*}{\cellcolor{color13TeV} 13}  &
			~\cite{ATLAS:2021upq} \\
			
			\hline\hline
	\end{tabular}}
	\caption{Relevant CMS and ATLAS searches for a heavy charged scalars.}
	\label{tab:CMS_ATLAS_heavy_charged_scalar}
\end{table}


\subsection{Other measurements}
\label{sec:other}


\begin{table}[hb!]
    \centering
    \begin{tabular}{|c|c|c|c|c|}
    \hline\hline
        Expt. & Observable & Value & $\sqrt s$ (TeV) & Ref.  \\
        \hline\hline
       \cellcolor{atlas}ATLAS & $\mathcal B(h\to \text{invisible})$& $<10.7\%$ &\cellcolor{color13TeV} 13 & \cite{ATLAS:2023tkt}\\
       \hline\hline
       \cellcolor{lep}LEP&\multirow{2}{*}{$\Gamma(Z\to \text{invisible})$}& $(0.499\pm0.0015)$ GeV& $0.088 - 0.094$&\cite{ALEPH:2005ab}\\
       \cline{1-1}\cline{3-5}
       \cellcolor{cms}CMS& & $(0.523\pm0.016)$ GeV& \cellcolor{color13TeV} $13$&\cite{CMS:2022ett}\\
        \hline\hline
       \cellcolor{lep}LEP&$\Gamma(W\to \text{invisible})$& $(0.032\pm0.060)$ GeV& $0.161-0.183$&\cite{ALEPH:1999jcv}\\
       \hline \hline
       \cellcolor{cms}CMS&$\Gamma(\text{top})$& $(1.36\pm0.127)$ GeV& \cellcolor{color8TeV} $8$&\cite{CMS:2014mxl}\\
    \hline\hline
    \end{tabular}
    \caption{Relevant branching fractions and decay widths of heavy SM particles~\cite{ParticleDataGroup:2024cfk}.}
    \label{tab:widths}
\end{table}


\begin{table}[hb!]
    \centering
    \begin{tabular}{|c|c|c|c|c|}
    \hline\hline
     Expt.&Process & Range (GeV) & $\sqrt s$ (TeV) & Ref.  \\
    \hline\hline
    \cellcolor{lep}LEP & $\tilde l^+ \tilde l^- \to l^+l^-+2\tilde\chi^0$ & $45 < M_{\tilde l}< 102$ & 0.183 -- 0.208 & \cite{LEP:susy}\\ \hline\hline
    \cellcolor{atlas}& $\tilde \ell^+ \tilde \ell^- \to \ell^+\ell^-+2\tilde\chi^0$ & $90<M_{\tilde \ell}< 350$ & \cellcolor{color13TeV} &  \cite{ATLAS:2022hbt}\\
    \cline{2-3}\cline{5-5}
    \cellcolor{atlas}\multirow{-2}{*}{ATLAS}& $\tilde \tau^+_L \tilde \tau^-_L \to \tau^+\tau^-+2\tilde\chi^0$&  $80<M_{\tilde \tau}< 450$ & \cellcolor{color13TeV} \multirow{-2}{*}{13}& \cite{ATLAS:2019gti} \\ \hline\hline
    \cellcolor{cms}& $\tilde \ell^+ \tilde \ell^- \to \ell^+\ell^-+2\tilde\chi^0$& $100<M_{\tilde \ell}< 800$ & \cellcolor{color13TeV} & \cite{CMS:2024gyw} \\
    \cline{2-3}\cline{5-5}
   \cellcolor{cms}\multirow{-2}{*}{CMS} & $\tilde \tau^+_L \tilde \tau^-_L \to \tau^+\tau^-+2\tilde\chi^0$ & $90<M_{\tilde \tau}< 500$ & \cellcolor{color13TeV} \multirow{-2}{*}{13}& \cite{CMS:2022syk} \\ \hline\hline
    \end{tabular}
    \caption{Slepton searches at the LEP and the LHC. Here, $l$ stands for three flavours of charged leptons whereas $\ell$ indicates electron or muon. These searches can be reinterpreted as charged Higgs searches.}
    \label{tab:SUSY_obs}
\end{table}


\FloatBarrier

\clearpage

\bibliographystyle{JHEP}
\bibliography{bib}{}
\end{document}